\documentclass[
reprint,
superscriptaddress,
notitlepage,
 amsmath,amssymb,
 aps,
floatfix,
bibliography
]{revtex4-1}
\usepackage{lipsum}
\usepackage{color}
\usepackage{ragged2e}
\usepackage{afterpage}

\usepackage{amsmath}
\usepackage[dvipsnames]{xcolor}

\usepackage{titlesec}

\titlespacing*{\section}
{0pt}{0.5cm}{0.5cm}

\titleformat*{\subsection}{\bf}{\raggedright}
\titlespacing*{\subsection}
{0cm}{0.5cm}{0.25cm}

\titleformat*{\subsubsection}{\it}{\raggedright}
\titlespacing*{\subsubsection}
{0cm}{0.5cm}{0.25cm}

\usepackage[paperwidth=210mm,paperheight=297mm,centering,hmargin=1.75cm,vmargin=2cm]{geometry}

\usepackage{eqnarray}

\usepackage{graphicx}
\usepackage{dcolumn}

\newcommand\asm[1]{}
\newcommand\green[1]{{#1}}
\usepackage{setspace}

\newcolumntype{C}[1]{>{\centering}m{#1}}

\setcitestyle{super}

\begin{document}

\title{\green{A Mechanical Route for Cooperative Transport in Autonomous Robotic Swarms}}

\author{Eden Arbel}
\affiliation{School of Physics and Astronomy, and the Center for Physics and Chemistry of Living Systems, Tel Aviv University, Tel Aviv 6997801, Israel}

\author{Luco L.K.M. Buise}
\affiliation{Department of Artificial Intelligence, Donders Center for Cognition, Radboud University, Nijmegen, Netherlands}

\author{Charlotte C.R.M.M. van Waes}
\affiliation{Department of Artificial Intelligence, Donders Center for Cognition, Radboud University, Nijmegen, Netherlands}

\author{Naomi Oppenheimer}

\author{Yoav Lahini}
\affiliation{School of Physics and Astronomy, and the Center for Physics and Chemistry of Living Systems, Tel Aviv University, Tel Aviv 6997801, Israel}

{}
\author{Matan Yah Ben Zion}
\email[]{matanbz@gmail.com}

\affiliation{School of Physics and Astronomy, and the Center for Physics and Chemistry of Living Systems, Tel Aviv University, Tel Aviv 6997801, Israel}
\affiliation{Department of Artificial Intelligence, Donders Center for Cognition, Radboud University, Nijmegen, Netherlands}




\begin{abstract}

Cooperative transport is a striking phenomenon where multiple agents join forces to transit a payload too heavy for the individual. While social animals such as ants are routinely observed to coordinate transport at scale,
 reproducing the effect in artificial swarms remains challenging, as it requires synchronization in a noisy many-body system.
 Here we show that cooperative transport spontaneously emerges in swarms of stochastic self-propelled robots. \green{Robots deprived of sensing and communication, are isotropically initialized around a passive circular payload, where directional motion is not expected without an external cue. And yet it moves.} We find that a minute modification to the mechanical design of the individual agent dramatically changes its alignment response to an external force. We then show experimentally that \green{by controlling the individual's friction and mass distribution}, a swarm of active particles \green{autonomously} cooperates in the directional transport of larger objects. Surprisingly, transport increases with increasing payload size\green{, and its persistence surpasses the persistence of the active particles by over an order of magnitude}. A mechanical, coarse-grained description reveals that force-alignment is intrinsic and captured by a signed, charge-like parameter with units of curvature. Numerical simulations of swarms of active particles with a negative active charge corroborate the experimental findings. We analytically derive a geometrical criterion for cooperative transport which results from a bifurcation in a non-linear dynamical system. Our findings generalize existing models of active particles
, offer new design rules for distributed robotic systems, and shed light on cooperation in natural swarms.
\end{abstract}

\maketitle
\newpage 

\section{Introduction}\vspace{-0.3cm}
Foraging ants teaming up to transport a large payload is a hallmark of agile cooperation in nature~\cite{Traniello1983,Franks1986,Czaczkes2013, McCreery2014,Buffin2018}. Groups of ants can forage synergetically, transporting items heavier than the summed capacity of the individuals~\cite{Franks1986}. 
The significance of cooperative transport in living systems, and the potential industrial applications of coordinating transport using simple agents with only local sensing attracted interest beyond entomology, inspiring researchers across disciplines \green{including swarm robotics~\cite{Brooks1989a,Kube1993,Cao1997,Chen2015,Alkilabi2017,Tuci2018,Savoie2019,Boudet2021,Giardina2022,Pratissoli2023}, non-equilibrium physics~\cite{Kaiser2013,Wensink2014,Bechinger2016,Li2019,Gompper2020,Granek2022}, and biology of social animals~\cite{Kube2000,Gelblum2015,Gelblum2016, Ron2018,Feinerman2018,Heckenthaler2023}}.



Recent research in collective behavior successfully captured emergent effects such as flocking or aggregation by treating individual agents as stochastic self-propelled particles with simple interaction rules~\cite{Vicsek1995,Tailleur2008,Bricard2013,BenZion2022}. This approach, however, proved limited in describing cooperative transport: a passive payload introduced to a swarm of active particles shows moderate, diffusive dynamics. Unless the payload has an explicit shape asymmetry, it only exhibits Brownian motion~\cite{Kaiser2013, Wensink2014, Bechinger2016, Gompper2020, Granek2022}.

Replacing the primitive active agents with robotic swarms augmented with sophisticated circuitry and advanced artificial intelligence also had limited outcomes without the aid of an external cue or manual positioning~\cite{Alkilabi2017}: robots with electronic feedback, proximity sensors, and communications, struggle to respond to the rapidly changing environment owing to frequent mechanical collisions of the robots with the payload and with one another~\cite{Kube2000, Chen2015, Tuci2018, Li2019, Savoie2019, Boudet2021, Giardina2022, Pratissoli2023}. \green{At the absence of an external cue, cooperative transport was restricted to small swarms, and required a manual pre-arrangement of the swarm relative to the payload~\cite{Tuci2018}.}
Programming robots to avoid mechanical collisions altogether suppressed the collective dynamics, leading many times to swarm-scale deadlocks~\cite{Rubenstein2014, Werfel2014}. 

\begin{figure*}[t]
    \centering
    \includegraphics[width=\linewidth]{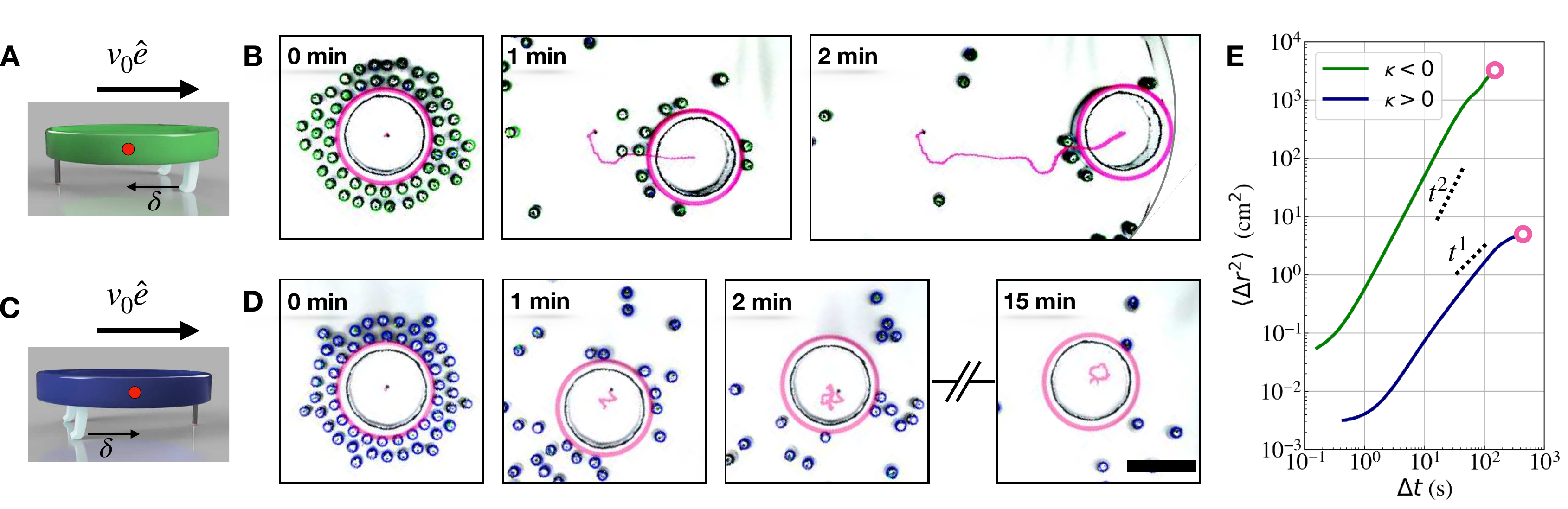}
    \caption{\footnotesize{{\bf Spontaneous cooperative transport of a payload by stochastic vibrational robots}. (A) A robot with a stiff rear leg and two soft front legs driven using vibration motors, and the center of mass (red dot) is behind the soft legs: $\delta<0$ (B) Time laps of a swarm of robots that spontaneously push a larger payload (diameter $2a = 28\rm{cm})$ moving it all the way to the arena's boundary. (C) A robot design with opposite leg polarity, where the stiff leg is in the front, and the soft legs are in the back (center of mass is forward of the soft legs $\delta>0$). (D) Robots with this design are deflected by the payload which shows only moderate, rather diffusive displacement. (E) Mean square displacement of the payload shows near ballistic motion ($\propto t^2$) for $\delta<0$ design (green), while with robots with $\delta>0$ (blue), the payload show orders of magnitude slower, near diffusive, motion ($\propto t^1$). Scale bar is 20 cm.}}
    \label{fig1}
    \vspace{-0.5cm}
\end{figure*}

In this work, we show that collective transport can spontaneously emerge in a swarm of rudimentary self-propelled particles, without any form of sensing, feedback, or control. This is achieved via a minor adjustment to the mechanical design of the self-propelled particle, which dramatically alters its orientation response to an external force. We show experimentally that it is possible to achieve negative force-alignment in which particles orient themselves opposite to the external force, and that a swarm of such robots cooperate in the directional transport of a larger, passive object (see Fig.~\ref{fig1} and Supporting Videos 1 and 2).
An inspection of the contact dynamics of an active particle with the ground (Supporting Videos 3 and 4) allows us to derive from first principles the pivotal contribution of mechanics to force-alignment  (Fig.~\ref{fig2}), thereby extending, and generalizing previous phenomenological descriptions for the equations of motion of self-propelled particles~\cite{Howse2007, Giomi2013, Lam2015, Deblais2018, Jin2019, Dauchot2019, Zhang2021, Boudet2021, Baconnier2022, Mirhosseini2022, Benzion2023}. 
We find that the force-alignment is intrinsic to active particles, and can be described by a signed, charge-like parameter with units of curvature, which we term ``curvity''.

We use this model in numerical simulations of stochastic active particles, where we observe cooperative transport when particles have a negative curvity, corroborating the experimental observations (see Fig.~\ref{fig3} and Supporting Videos 1,2,5, and 6). Surprisingly, in both experiments and simulations the transport propensity \textit{increases} with increasing payload size (see Fig.~\ref{fig4}). We analytically derive a condition for transport, which depends on the geometrical curvature of the payload as well as the intrinsic curvity of self-propelled particles. The condition is consistent in both simulations and experiments, offering a geometrical criterion for cooperative transport.
	

\section{Experiments in cooperative transport}\vspace{-0.3cm}	
Stochastic self-propelled robots were built following a modified bristle bot design~\cite{Giomi2013,Deblais2018,Dauchot2019,Benzion2023}. A robot (sizing 5-6 cm in diameter) is driven by two vibration motors mounted on a tripod with one stiff leg and a pair of asymmetric soft legs (see Fig. \ref{fig1} and Supplementary Information Section~\ref{secExperiments}). Vibrations induce noisy forward motion which defines the robot's heading, $\hat{e}$ (see Figs.~\ref{fig1}A, C). In the experiments, a large circular passive payload was placed in the middle of a symmetrical arrangement of robots, which were then turned on, setting the swarm into motion. With the traditional design (soft legs placed at the back), robots sporadically push the passive payload which exhibits Brownian-like motion --- with each collision, robots turn away from the payload (Fig. \ref{fig1}C, D, E,  and Supporting Video 5). In contrast, when the soft legs are placed at the front, the swarm spontaneously breaks symmetry and propels the payload in a near ballistic trajectory (see Fig.~\ref{fig1}A, B, E and Supporting Video 1). Here, with each collision, robots tend to turn into the payload and progressively push it until reaching the perimeter of the arena (150 cm diameter). \green{The cooperative transport emerges autonomously, and does not require an external, directional cue, nor manual pre-arrangement}. We further find the effect to increase with payload diameter ($2a = 7-32\; \rm{cm}$), swarm size ($N=1-53$ robots) in both a custom-made and a modified commercial multi-robot platform~\cite{Rubenstein2014,Benzion2023} (see Fig.~\ref{fig4} and Supporting Information Section~\ref{secSwarmSize}).

High-speed video imaging offered insight into the origin of force-alignment (see Fig.~\ref{fig2}, Supporting Videos 3, 4, and Supporting Information Section~\ref{secHighSpeedImaging}). While moving, the robot's stiff and soft legs interact differently with the ground --- the stiff leg has higher restitution, and spends a longer duration in the air, whereas the softer legs show only moderate hopping. This difference leads to a differential fore-aft friction which lies at the heart of the mechanical origin of the force-alignment --- robots with soft legs at the back align with an external force (descend downhill), whereas robots with soft legs at the front align against the force (ascend uphill). The difference between the two designs is revealed in the presence of an external force or boundaries. In the absence of such,  their dynamics are qualitatively indistinguishable.

Force-alignment is generic to self-propelled particles regardless of the locomotion mechanism and should be expected in general both on the macroscopic and the microscopic scale. In the next section, we derive the mechanical origin of force-alignment from first principles.

\section{Mechanical origin of signed force alignment}\vspace{-0.3cm}
\begin{figure*}[t]
    \centering
    \includegraphics[width=0.75\linewidth]{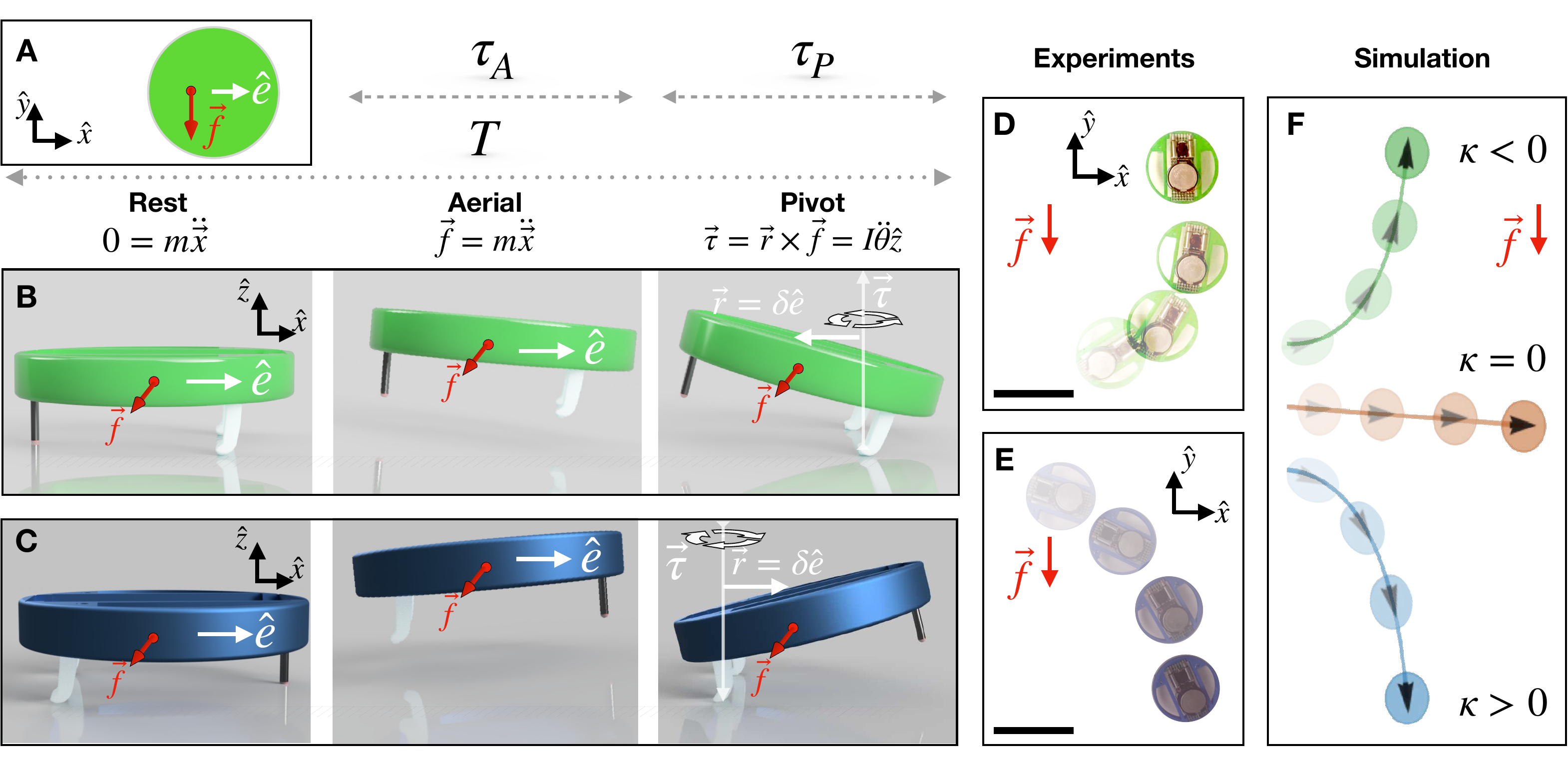}
    \caption{\footnotesize{{\bf The mechanical origin of force-alignment stands on the fore-aft restitution difference}. (A) A top view of an active particle self-propelled along its heading ($\hat{e}$) subjected to an in-plane external body force ($\vec{f}$) acting on the center of mass (CoM, red dot). (B) The quasi-two-dimensional motion of a bristle bot has three characteristic phases: (I) At {\it Rest} all legs are on the ground and the external force is balanced by static friction, the robot does not move. (II) In the {\it Aerial} phase, the robot is completely aloft with constant linear acceleration along the external force. (III) Having softer legs creates a {\it Pivot} phase, where the robot is partially touching the ground, and the external force creates a torque around the pivot axis. The softer legs are in front of the CoM, and the robot turns {\it against} an external force (see also Supporting Video 3). (C) A robot with a stiffer front leg ($\delta>0$, soft legs at the back) goes through a similar sequence but rotates in the opposite direction, i.e. {\it along} the external force (see also Supporting Video 4). (D) A robot with soft front legs goes against an external force and climbs up an inclined plane. (E) A robot with soft rear legs goes down an inclined plane (see also Supporting Video 4). (F) Trajectories of numerical simulation of Eqs.~\ref{eqForce},~\ref{eqAlignment} show that particles with negative curvity ($\kappa <0$) turn and move against an external force, like robots with soft front legs (green), and particles with positive curvity ($\kappa>0$), turn along the external force, like robots with soft rear legs (blue). A particle with a theoretical zero curvity (like ABP), drifts along the external force, but does not reorient its heading. Scale bars are $10\;\rm{cm}$.}}
    \label{fig2}
    \vspace{-0.5cm}
\end{figure*}

The microscopic origin of force-alignment is revealed by considering the instantaneous acceleration, $\vec{a}$, of a vibrationally propelled robot under an external body force, $\vec{f}$, acting in the plane of motion.  \green{Below we outline the coarse-graining of the rapid hopping dynamics, and derive effective equations of motion of a granular active particle dominated by inertia and dry friction. We will reproduce previous work that assumed overdamped dynamics, and force alignment based on symmetry~\cite{Lam2015}, however we will show that their effective parameters (mobility and force-alignment) are controlled by inertial quantities (mass and moment of inertia).}

We consider the motion to have three characteristic phases: I --- rest, II --- aerial, and III --- pivot, with a mean overall duration $T$ (see Fig.~\ref{fig2}). A robot starts at {\it rest} (I) with all contact points on the ground, thereby the external force is perfectly balanced by static friction and there is no motion ($m\vec{a} = 0$). The robot then jumps forward with a horizontal speed of $\vec{v} = v_A \hat{e}$ having a typical time aloft of $\tau_A$. While in the {\it aerial} phase (II) the robot accelerates by the external force $m\vec{a} = \vec{f}$. For simplicity, we treat contact with the substrate as having perfect static friction, and when the robot lands it loses all its momentum. Combining phases I and II results in a coarse-grained velocity proportional to the sum of active velocity and the external force (Eq. \ref{eqForce}), where the nominal speed is $v_0\equiv \frac{v_A\tau_A}{T}$ and the mobility is $\mu \equiv \frac{\tau_A^2}{2mT}$. This formalism is similar to Drude's model that leads to linear Ohm's conductivity where charge carriers in a conductor lose their momentum during collisions. 
Inevitably, contact friction is not equal on all legs, and empirically we find that the robot spends a longer duration, $\tau_P$, on the softer legs acting as a pivot. During the {\it pivot} phase (III), static friction with the contacting foot restricts linear motion ($m\vec{a} \approx 0$), however, the robot can rotate, as it experiences a torque $\vec{\tau}$. The torque is the result of the external force acting on the center of mass which in general is displaced from the axis of rotation, $I \vec{\alpha} = \vec{\tau} = \delta\hat{e}\times\vec{f}$, where $I$ is the moment of inertia around the rotation axis, and the lever arm, $\delta$, is the offset of the center of mass from the axis along the orientation vector $\hat{e}$ (see Fig.~\ref{fig2}B, C). The offset, $\delta$, can be positive or negative, respectively resulting in positive or negative force-alignment. 

Phase III gives the microscopic basis for force-alignment on which our model rests. Combining the instantaneous dynamics of phases I-III results in coarse-grained equations of motion 

\vspace{-0.5cm}
\begin{equation}
    \frac{d}{dt}\vec{r} \equiv \vec{v} = v_0 \hat{e} + \mu \vec{f}
    \label{eqForce}
\end{equation}

\vspace{-0.5cm}
\begin{equation}
    \frac{d}{dt} \hat{e} = \kappa \hat{e} \times \left(\vec{v}\times\hat{e}\right),
    \label{eqAlignment}
\end{equation}
where $\kappa \equiv \frac{m\delta}{I}\left(\frac{\tau_P}{\tau_A}\right)^2$ acts as an effective charge-like parameter of an active particle and is a key result of our model (for details see SI Section~\ref{secMechanics}). We term $\kappa\;$ {\bf curvity}, as it has units of curvature, stemming from the particle's activity. 
Similarly to an electric charge, $\kappa$ is signed, and its sign is controlled by an internal symmetry.  The sign and magnitude of the curvity follow $\delta$, the signed offset of the center of mass.

\green{Remarkably, despite not having any formal viscus drag, Eq.~\ref{eqForce} has the same structure as the equations used to describe drag-dominated micro-swimmers acting in the low Reynolds number regime~\cite{Gompper2020}, where inertial quantities are justifiably neglected, and velocity is proportional to the external force through a mobility constant ($\mu$). Previous work on granular active matter already assumed that dry macroscopic objects can be described using overdamped dynamics~\cite{Lam2015,Dauchot2019}. The derivation above shows that while the equations of motion take the same structure, they are controlled by effective parameters that directly depend on inertial quantities, like mass ($m$) and moment of inertia ($I$).} Equations~\ref{eqForce} is also found in the extensively used model of Active Brownian Particles (ABP)~\cite{Howse2007,Tailleur2008,Marchetti2013,Redner2013,Fily2014,Fily2015,Solon2015}. Equation~\ref{eqAlignment} accounts for the contribution of Force-Alignment in an Active Brownian Particle (FAABP). The equations only require the mean value of the different parameters, with no particular significance to the order of the three phases.

\begin{figure*}[t]
    \centering
    \includegraphics[width=\linewidth]{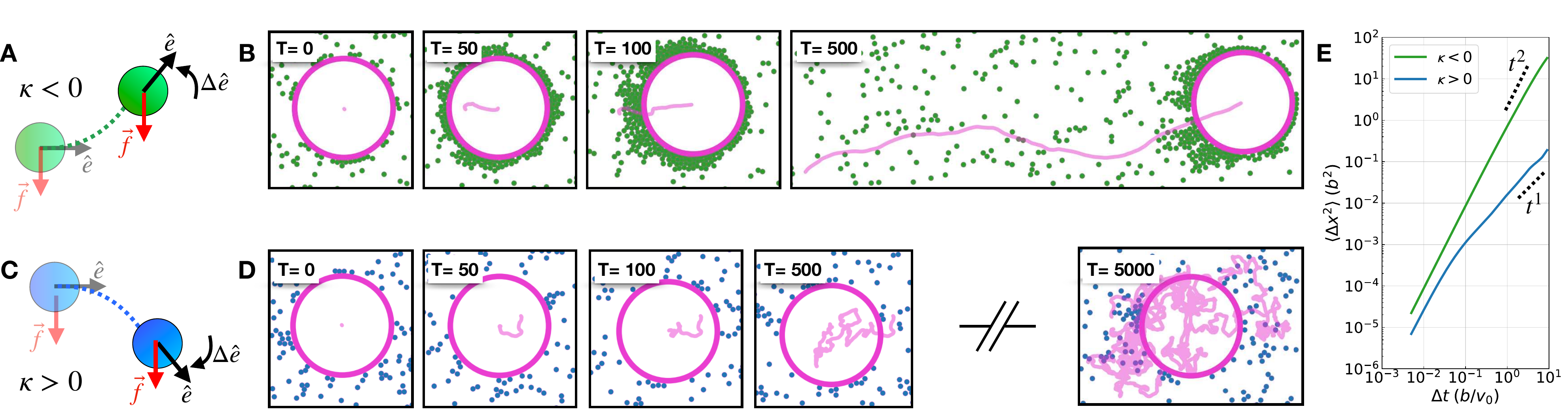}
    \caption{\footnotesize{{\bf Numerical simulations of self-propelled particles with negative force-alignment show cooperative transport.} (A) Schematic of the motion of a self-propelled particle with negative curvity ($\kappa<0$) that turns its heading ($\hat{e}$) against an external force. (B) Time sequence from a simulation of 1000 particles shows a progressive accumulation of the particles on one side of the payload followed by its transport. An active wake is formed at the rear of the payload and continually exchanges particles with the surroundings in a dynamic steady-state (see also Supporting Video 2). (C) Schematic of the motion of a self-propelled particle with positive curvity, that turns its heading along an external force. (D) Time sequence of a payload with one thousand active particles with positive curvity shows a diffusive trajectory (see Supporting Video 6). (E) Mean square displacement of the payload's trajectory shows near ballistic motion ($\propto t^2$) when $\kappa<0$ (green) but near diffusive motion ($\propto t^1$) when $\kappa>0$ (blue).}}
    \label{fig3}
\vspace{-0.25cm}
\end{figure*}

An important consequence of the microscopic model presented here is to identify that $\kappa$ is signed and to offer a powerful design rule. For example, in the point mass limit ($I=m\delta^2$), the curvity is inversely proportional to the offset $\kappa \propto 1/ \delta$, and when the offset is negative (center of mass is behind the soft legs, $\delta<0$), robots turn against an external force. \green{The curvity can be also computed for an arbitrary shape (e.g. rod-like) and mass distribution, by evaluating the robot's moment of inertia relative to the pivot axis, as the balance of the increased lever arm ($\kappa \propto \delta$), with the increased moment of inertia ($\kappa \propto 1/I$). The derivation above also shows that the effective parameters ($v_0$, $\mu$, and $\kappa$) are not independent of one another, and how they are linked by the robot's inertial properties. It is interesting to note that previous work on E.coli~\cite{Lauga2006} and Paramecium~\cite{Roberts2010} showed that even on the micro-scale, an increase in the size of a micro-swimmer leads to an increase in the radius of curvature of its trajectory, suggesting a potential extension of the collective dynamics described here to the domain of micro-swimmers~\cite{Gompper2020}.}


\section{Cooperative transport in numerical simulations}\vspace{-0.3cm}
We tested numerically swarms of FAABPs by adding orientational noise to Eq.~\ref{eqAlignment} and short-range repulsion, in a simulation engine using 5th-order Runge-Kutta integration. The orientational noise has zero mean, $\langle \vec{\xi}\rangle = 0$, with a Gaussian distribution of width $\langle \xi^2\rangle = 2\Delta t k_BT$ ($\Delta t$ is the simulation time step, and $k_BT$ sets the magnitude of the noise), with particles modeled as soft discs of radius $b$ (see Supporting Information Section ~\ref{secSimulations} for details).

\subsubsection*{Individual Force-Aligning Active Brownian Particles}

 Simulated dynamics of individual FAABPs with a constant force reproduce experimental trajectories of robots moving on an inclined plane (see Figs.~\ref{fig2}D, E, F). FAABPs with a positive curvity ($\kappa >0$) turn in the direction of the force similarly to robots having their center of mass in front of their soft legs ($\delta>0$), whereas FAABPs with a negative curvity ($\kappa < 0$) turn anti-parallel and move \textit{against} the external force, like the robots with their center of mass behind the soft legs ($\delta <0$). The singular, zero curvity FAABP ($\kappa=0$), is simply an ABP --- its heading is unaffected as it drifts in the direction of the external force (Fig.~\ref{fig2}F).

\subsubsection*{Cooperative Transport in Swarms of Force-Aligning Active Brownian Particles}

\begin{figure}[t]
    \centering
    \includegraphics[width=\linewidth]{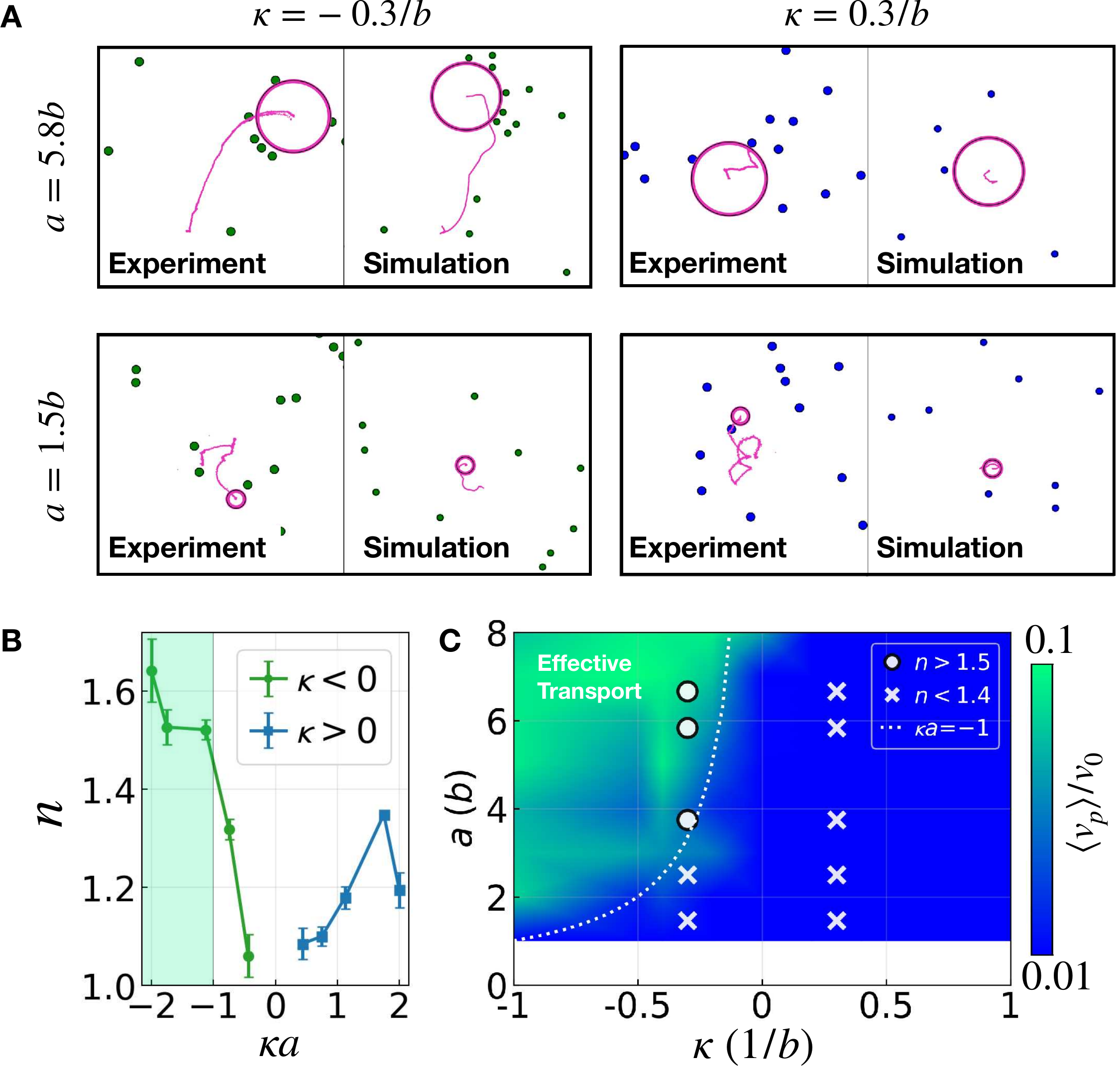}

  \caption{\footnotesize{{\bf Larger payloads are better transported.} (A) Individual trajectories become increasingly persistent for larger payloads and negative curvity in both experiments and simulations. (B) The power law ($n$) of the mean square displacement is closer to ballistic motion ($n>1.5$), when $\kappa a <-1$, and closer to diffusive ($n<1.4$), when $\kappa a > -1$. Each point is the average $n$ of four experiments with error bars showing standard error (see Supplementary Information Section~\ref{secExperiments} for details). (C) Simulations of a payload of radius $a$, in a swarm of 200 FAABPs of curvity $\kappa$, moves an order of magnitude faster when $\kappa a <-1$ (simulation results measure mean speed of payload $\langle v_p\rangle$ relative to nominal speed of FAABPs, $v_0$, see Supporting Information Section~\ref{secSimulations} for details). Circles show experimental results with near ballistic MSD ($n>1.5$), and $\times$ denotes experiments where power law is closer to diffusive ($n<1.4$). Dashed line follows the analytical prediction for cooperative transport ($\kappa a = -1$, see Eq.~\ref{eqTransport}), and found at the boundary between the two dynamical regimes.}}
    \label{fig4}
    \vspace{-0.25cm}
\end{figure}

We tested the effect of a passive particle of radius $a$ on a randomly distributed swarm of FAABPs ranging in sizes between $N \in [1,1000]$, with negative curvities, $\kappa <0$. We observe that after a short transient where the swarm homogeneously accumulates around the passive particle, symmetry is spontaneously broken and particles form an active wake on one side that propels the passive payload (see Fig.~\ref{fig3} and Supporting Video 2). \green{In line with the experimental findings, the transport emerges autonomously, despite the initial isotropic random arrangement of the active particles}. The passive particle shows elongated trajectories, larger than its size, and larger than the simulation box (for periodic boundary conditions). The direction of transport is different from one run to the other, and the active wake is in a dynamic steady state, constantly exchanging the participating FAABPs. A similar effect is also observed in the non-periodic simulation, excluding the effect of the boundary. Transport is also observed when FAABPs are non-interacting (can pass through one another but not through the passive particle) excluding the role of Motility Induced Phase Transition~\cite{Tailleur2008,Cates2015}. \green{This means that cooperation emerges not by direct robot-to-robot interaction, but instead by a proxy --- the passive payload.} Transport is completely absent for FAABPs with positive curvity ($\kappa > 0 $) or for smaller payloads, where the passive particle only shows a diffusive trajectory (see Fig.~\ref{fig3}C, D, Fig.~\ref{fig4}, and Supporting Video 6).

\section{Dependence on payload size}\vspace{-0.3cm}
Counter-intuitively, cooperative transport is enhanced with increasing payload radius, $a$. \green{By contrast to thermal fluctuations, where a particle's diffusion decreases with its size~\cite{Einstein1905}, here we find that both the payload's speed ($v_p$) and persistence length ($l_p$) increase with its size, with an overall  increased long term effective diffusion ($\propto v_p l_p$)~\cite{Howse2007}.} Performing 116 experiments and over 1000 simulations (varying payload size, curvity, robot count, and orientational noise, see Supporting Information), we found that in both experiments and simulations, larger payloads are better transported, provided that the curvity of the active particles is sufficiently negative (see Fig.~\ref{fig4}). \green{In the experiments, the payload's weight is proportional to the payload's radius ($m\propto a$, see SA \ref{secPayloadSize}), for a combined super-linear increase in mass transport.} Trajectories of a larger payload in a swarm of robots with negative curvity show more persistent motion compared to that when the curvity is positive or the payload is smaller (Fig.~\ref{fig4}A). Experimentally, there is a considerable increase in the average power-law of the mean square displacement of the payload when $\kappa a <-1$ (see Fig.~\ref{fig4}B).  Exploring the $\kappa$-$a$ phase space in simulation reveals two phases, with an order of magnitude increase in the mean speed of the payload (Fig. \ref{fig4}C). The phase boundary for both experiments and simulations lies at $\kappa a = -1$.

Finding a consistent condition for a far-from-equilibrium many-body process, in both experiment and simulation is quite striking, especially in lieu of the profoundly different physical laws governing the two: Simulations model particles in the over-damped limit, with drag proportional to the particle's diameter, the self-propulsion is smooth,  the orientational noise is Gaussian, and the motion is strictly two-dimensional. By contrast, robots in experiments are inertial objects, making an effective active granular gas, with solid friction with the ground and with one another, an intermittent vibrational self-propulsion, a non-Gaussian orientational noise, and their rapid hopping is quasi-two-dimensional. Having such deprecate physical dynamics lead to an identical condition for cooperative transport suggests a deeper explanation. 
We next show that the condition for cooperative transport is geometrical, stemming from the interplay of the active particles' curvity, $\kappa$, and the curvature of the circular payload, $1/a$. 


\section{Geometric criterion for cooperative transport}\vspace{-0.3cm}

\begin{figure}[b]
    \centering
    \includegraphics[width=\linewidth]{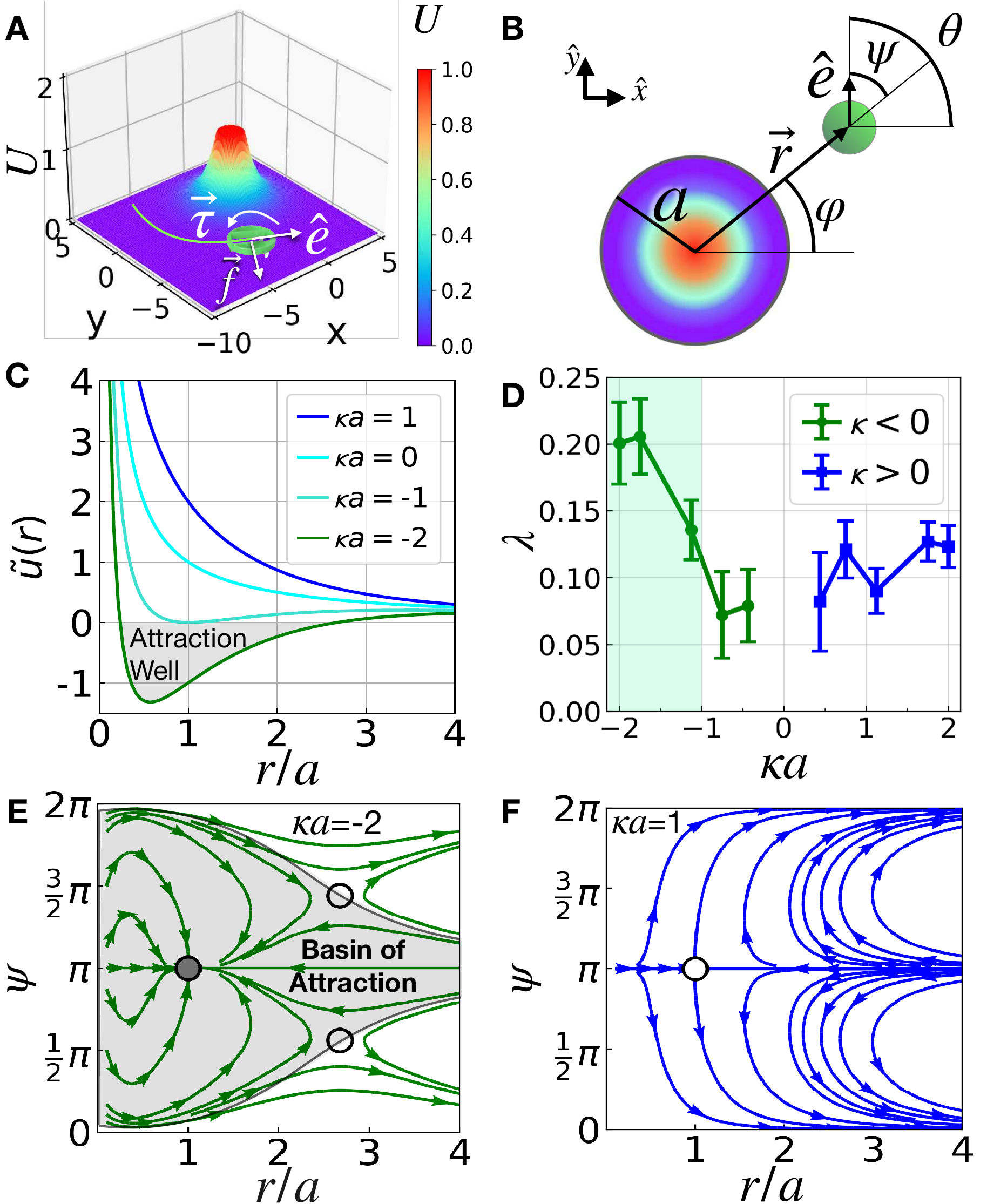}
    \caption{\footnotesize{{\bf FAABPs with a negative curvity show an effective attraction to a repulsive potential.} (A) Illustration of an active particle near a repulsive potential. (B) Configurational coordinates [position $\vec{r}=\left(r,\varphi\right)$ and heading $\theta$] of an active particle (green) near a circular repulsive potential. $\psi$ is the angle of the heading relative to the potential center. (C) An effective attraction well is formed when  $\kappa a <-1$.  $\tilde{u}\left(r\right)$ is the prefactor in Eq.~\ref{eqPsi}, and in regions where $\tilde{u}\left(r\right)<0$, self-propelled particles are effectively attracted to an otherwise repulsive potential. (D) Experiments where $\kappa a <-1$ show an effective attraction as an increase in the mean linear density, $\lambda$, of robots at the perimeter of the payload. Each point is the average $\lambda$ of 4 runs with standard error. (E) Phase portraits of Eqs.~\ref{eqRadial} and ~\ref{eqPsi} in the distance and relative orientation plane ($r$-$\psi$) display a basin of attraction (gray region) when $\kappa a =-2$: an active particle is effectively attracted to a repulsive potential. At the linearly stable fixed point ($r=a$, $\psi = \pi$, filled circle) self-propulsion is balanced by the repulsive force.  (F) When $\kappa a = 1$, the fixed point becomes a saddle (empty circle), and there is no activity-induced attraction.}}
    \label{fig5}
    \vspace{-0.25cm}
\end{figure}

We start by modeling a circular payload as a repulsive, two-dimensional, radially symmetric potential fixed at the origin, $U=U\left(r\right)$, exerting a repulsive force on the active particles: $\vec{f} = -\vec{\nabla} U (r) = \Gamma \left(r\right)v_0/\mu \hat{r}$ (see  Fig.~\ref{fig5}A). Active particles interact with this circular obstacle following Eqs.~\ref{eqForce} and~\ref{eqAlignment}. $\Gamma(r)$ sets the radial profile of the force and is chosen such that the magnitude of the repulsive force at the payload's perimeter exactly balances the nominal speed of the active particle, $\Gamma(r=a) = 1$, effectively setting the payload's size. Inspired by previous work on repulsive particles~\cite{BenZion2024}, keeping the potential profile implicit (exponential decay, soft-core, \green{Yukawa}, etc.), makes the result below more general. Circular self-propelled particles in 2D have three degrees of freedom (see Fig.~\ref{fig5}B): a radial and azimuthal position $\left(r,\varphi\right)$, and a heading relative to the x-axis, $\hat{e} \equiv \left(\rm{cos}\theta, \rm{sin}\theta\right)$. The system has rotational symmetry and the dynamics depend only on the orientation of the heading relative to the center of the potential $\psi \equiv \theta - \varphi$. Plugging the radial force term into the FAABPs' equations of motion (Eqs.~\ref{eqForce},~\ref{eqAlignment}) gives a dynamical system described by two non-linear coupled first-order differential equations:
\begin{equation}
    \dot{r} =   v_0 \left[\rm{cos} \psi + \Gamma (r)\right]
    \label{eqRadial}
\end{equation}

\vspace{-0.5cm}
\begin{equation}
    \dot\psi = -v_0\left[\kappa \Gamma (r) +\frac{1}{r}\right] \rm{sin}\psi
    \label{eqPsi}
\end{equation}
(see Supplementary Information section \ref{secDynamics} for a detailed derivation). At $\psi = 0$ the active particle points away from the payload, and at $\psi = \pi$, it fronts the payload. When the prefactor in Eq. \ref{eqPsi}  switches sign ($\tilde{u}(r) \equiv \kappa\Gamma(r)+1/r=0$), an active particle is effectively attracted to the repulsive potential (see Fig.~\ref{fig5}C). Given the definition of $\Gamma\left(r\right)$, this can be satisfied when
\begin{equation}
    \kappa + 1/a < 0.
    \label{eqTransport}
\end{equation}

The condition in Eq.~\ref{eqTransport} presents a geometrical criterion for cooperative transport: once the curvity is sufficiently negative, instead of being scattered away ($\psi\rightarrow 0$), an active particle colliding with the obstacle re-orients sufficiently fast into the receding repulsive hill to continually push against the obstacle ($\psi\rightarrow\pi$). \green{The inequality in Eq.~\ref{eqTransport} is agnostic to whether the curvity or the curvature is negative, and could be applied more generally. Even if the force alignment is non-negative (positive or zero), a self-propelled particle could display effective attraction to a concave boundary, provided that its curvature is sufficiently negative ($1/a<0$). This has been previously observed in self-propelled particles interacting with a concave obstacle~\cite{Kaiser2013}, in a single confined active particle interacting with the inner concave walls of a harmonic trap~\cite{Dauchot2019}, and more recently, in active particles interacting with one another~\cite{Casiulis2024}.}

\begin{figure}[t]
    \centering
    \includegraphics[width=\linewidth]{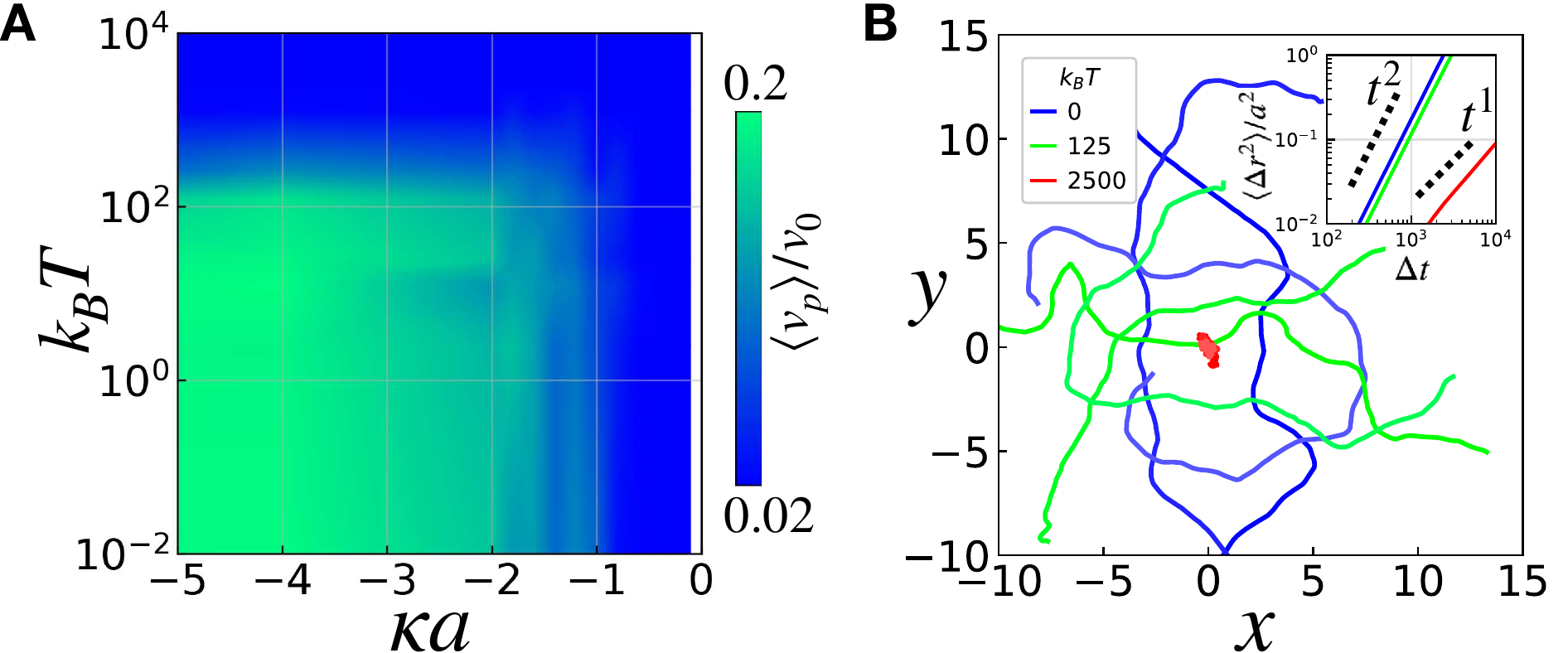}
    \caption{\footnotesize{{\bf Cooperative transport is robust to a range of re-orientational noise of the active particles.} (A) Phase diagram of the mean speed of the payload shows an order of magnitude increase when the condition in Eq.~\ref{eqTransport} is met, even at orientational noises above $100 k_BT$ (for details see Supplementary Information Section \ref{secSimulations}). (B) Individual trajectories of the payload and the ensemble averages of the mean square displacement (inset) for the case of  $\kappa a = -4$.  At low orientational noise (up to $125k_BT$) the passive payload moves in extended trajectories  (blue, green) with near ballistic MSD ($\propto t^2$, inset). At sufficiently high orientational noise ($k_BT = 2500$), the payload shows only modest motion (red blob) with near diffusive MSD ($\propto t^1$, inset). }}
    \label{fig6}
    \vspace{-0.25cm}
\end{figure}

Phase portraits of the dynamical systems above ($\kappa a = 1$) and below ($\kappa a = -2$) the transition, show a local topological change at the fixed point where the active particle is pushing against the payload, $r/a=1$, $\psi = \pi$ (see Fig.~\ref{fig5}E, F). When Eq.~\ref{eqTransport} is satisfied, the dynamical system undergoes a bifurcation, and the saddle point turns into a linearly stable sink that attracts active particles (see Supplementary Information Section ~\ref{secLinearStability}). In both experiments and simulations, this effective attraction is manifested in an enhanced kissing number $N_{\rm{kiss}}$ of robots touching the payload (see Figs.~\ref{fig1},~\ref{fig3},~\ref{fig4} and Supporting Videos 1, 2 and 5, 6), with an increased linear filling fraction, $\lambda\equiv N_{\rm{kiss}} b/a$ (see Fig.~\ref{fig5}D). In a system that meets the condition in Eq.~\ref{eqTransport}, the effective attraction and the resulting cooperative transport are robust over a range of orientational noises (see Fig.~\ref{fig6}).

\begin{figure}[t]
    \centering
    \includegraphics[width=\linewidth]{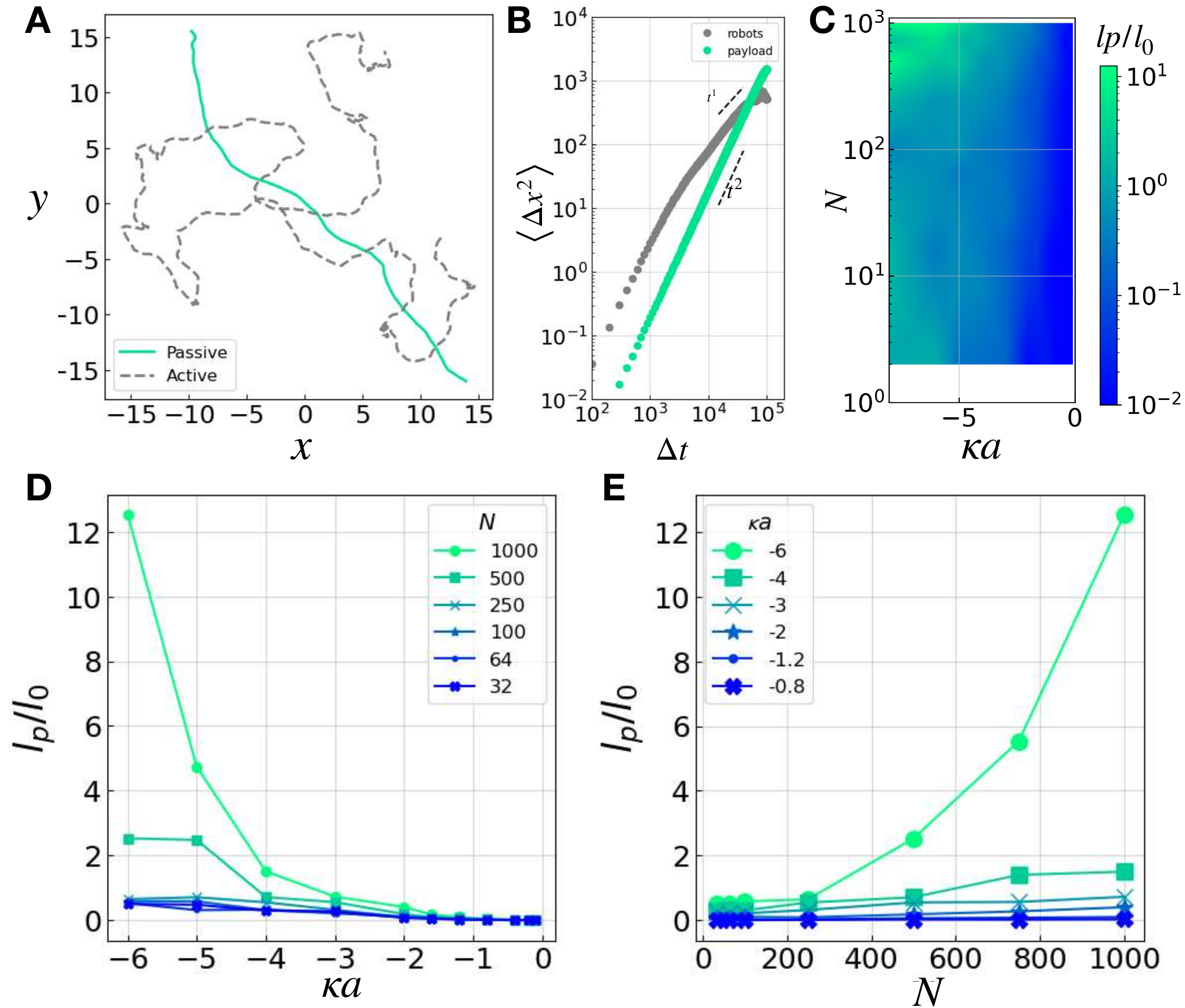}
    \caption{\footnotesize{\green{{\bf The cooperative transport is enhanced with both payload and swarm size}. (A) The trajectories and (B) MSDs of a simulation of $N=1000$ active particles (gray) and one passive payload (green, $\kappa a = -6$) show that the payload moves more slowly (shorter trajectory, lower MSD) but is more persistent ($\rm{MSD}\propto t^2$) than the active particles. (C) The persistence of a passive payload is amplified by over an order of magnitude relative to the active particles ($l_p/l_0>10$) with increasing interaction strength ($\kappa a$) and swarm size ($N$). Sections of the $\kappa a - N$ phase space show a sharper increase in the amplification with increasing swarm size (D), and a super linear marginal contribution of swarm size to the persistence length of the passive particle (E), expected from a cooperative system.}}}
    \label{fig7}
    \vspace{-0.25cm}
\end{figure}

\green{
\section{The cooperative nature of the transport}\vspace{-0.3cm}
We find that the persistence of the payload ($l_p$) increases with the number of active particles, and can even surpass the persistence of the active particles themselves ($l_0$, Fig.~\ref{fig7}A,B). This effect becomes clear when measuring the amplification of the persistence length ($l_p/l_0$): with increasing interaction ($\kappa a$ more negative), the amplification increases faster with increasing swarm size (Fig.~\ref{fig7}D). Moreover, at a given interaction strength, the amplified persistence increases super-linearly with the number of robots ($N$), a hallmark of a cooperative swarm~\cite{Hamann2018}. Overall, both the speed (Figs.~\ref{fig4},~\ref{fig6}), and the persistence length (Fig.~\ref{fig7}) of the payload increase with its size ($a$). The effective equations of motions derived above can explain this pronounced effect.}

Once the payload starts moving, the dynamics are no longer isotropic. A fluctuation driving the passive particle to the right, $u\hat{x}$ (w.l.o.g.), spontaneously breaks symmetry and introduces an explicit dependence on the azimuthal coordinate in Eqs.~\ref{eqRadial} and~\ref{eqPsi}, as well as an additional dynamical equation for the azimuth itself: $\dot{\varphi} = \left(v_0 {\rm sin} \psi + u \,\rm{sin}\varphi\right)/r$, leading to a dynamical system with three variables:
\begin{eqnarray}
&&\dot{r} =   v_0 \left[\cos \psi + \Gamma (r)\right] - u \cos\varphi
\label{eqRadialMoving}\\
&&\dot\psi = -v_0\left[\kappa \Gamma (r) +\frac{1}{r}\right] \sin\psi - \frac{u}{r}\sin\varphi
\label{eqPsiMoving}\\
&&\dot\varphi = \frac{v_0}{r}\sin\psi + \frac{u}{r}\sin\varphi,
\label{eqPhiMoving}
\end{eqnarray}

(see SI ~\ref{secMovingObject} for derivation). In the isotropic case (Eqs.~\ref{eqRadial} and~\ref{eqPsi}), there was a fixed point for any combination of the heading ($\theta$) and azimuth ($\varphi$) that satisfy:  $\psi \equiv \theta - \varphi = \pi$. When the payload is already moving, this is no longer the case. There are only two fixed points for the azimuthal degree of freedom, either  pushing against the payload's motion ($\varphi = 0$, unstable) or along its motion ($\varphi= \pi$, stable).  This means that when a group is transiting a payload, it preferentially recruits further individuals to push in the same direction, facilitating cooperative transport.



\section{Conclusions}\vspace{-0.3cm}
In this work, we found that a mechanical response to external forces alone can directly lead to cooperative transport. \green{We showed that transport emerges spontaneously and autonomously in a swarm of stochastic, self-propelled particles with no explicit sensing or decision-making, nor external cue whatsoever, important in the growing field of multi-robot systems}. We identified the key role of force-alignment response and traced its mechanical origin by coarse-graining the equations of motion from first principles. We found that an intrinsic parameter, which we term curvity, controls the sign and magnitude at which the orientation of an active particle responds to an external force, thereby setting the characteristic curvature of its trajectory. We discovered that particles with negative curvity tend to turn against an external force and push against obstacles. We experimentally fabricated such particles and presented a mechanical design rule for their construction, offering a route for engineering cooperative transport.

We then compared experiments and simulations over a range of parameters (including payload size, swarm size, curvity, and system noise) finding a consistent criterion for the emergence of cooperative transport as spontaneous symmetry breaking. The criterion is given by geometrical quantities, where self-propelled particles can become attracted to an otherwise repulsive potential. The criterion shares a mathematical structure with the Young-Laplace equation~\cite{DeGennes2002}, where the stability of a three-dimensional fluid interface is conditioned by two local curvatures, suggesting a link between interfacial phenomena, boundaries, and active matter.

Being of a geometrical origin, force-alignment can be tuned on the micron-scale: analogous dynamics were also observed in bacteria~\cite{Lauga2006}. A marriage of force-alignment with emerging colloidal technologies of artificial microswimmers harbors a potential for designing cooperative transport at the cellular level~\cite{Maass2016,Gompper2020,BenZion2022}. 
Moreover, effective attraction and repulsion can be further tuned by designing non-circular payloads with variable curvature (positive and negative) in tandem with the curvity of the self-propelled particles. Tunable attraction and repulsion, combined with cooperative transport, offer an activity-based architecture for unlocking new paradigms in modeling, as well as programming, far-from-equilibrium self-assembly.

Finally, although foraging ants are not simple stochastic particles, and make complex decisions based on sensory information, our findings suggest an underlying mechanism for scalable cooperation in nature. 





\vspace{0.25cm}

{\bf Supporting videos}

\begin{itemize}
    \setlength{\itemsep}{1pt}
    \setlength{\parskip}{0pt}
    \setlength{\parsep}{0pt}
    
    \item[] Video 1 --- Cooperative Transport - Experiment (53 robots)

    \item[] Video 2 --- Cooperative Transport - Simulation (1000 particles)

    \item[] Video 3 --- High Speed Imaging Negative Force-Alignment - Ascending

    \item[] Video 4 --- High Speed Imaging Positive Force-Alignment - Descending

    \item[] Video 5 --- Diffusive Motion of a Passive Payload - Experiment (53 robots)

    \item[] Video 6 --- Diffusive Motion of a Passive Payload - Simulation (1000 particles)
\end{itemize}
\bibliography{cooperativeTransport}
\vspace{-0.4cm}
\section*{Acknowledgment}
\vspace{-0.25cm}
We acknowledge I. Kolvin, C. Kelleher, and O. Dauchot for critical reading of the manuscript, and Y. Roichman for supplying Kilobots. This work was supported in part by the Israel Science Foundation grants 2096/18 and 2117/22 and the Israeli Ministry of Aliya, and by the project Dutch Brain Interface Initiative (DBI2), project number 024.005.022 of the research programme Gravitation financed by the Dutch Research Council (NWO).

\clearpage

\onecolumngrid

\newgeometry{top=25mm, bottom=25mm, right=25mm, left=25mm}

\doublespacing

\begin{center}
{\bf A Mechanical Route for Cooperative Transport in Autonomous Robotic Swarms --- Supporting Information}\label{secMethods}
\end{center}

\section{Experimental Supporting Information}\label{secExperiments}
	
\subsection{Robots Design and properties}

Four different self-propelled robots of two types were used in his work: {\it Custom Built Robots} and {\it Kilobot Based Robots}. Below is a detailed description of their construction, modeling, and mechanical properties. Table ~\ref{tblRobots} summarizes the mechanical properties of the robots.

\subsubsection{Custom Built Robots}
\vspace{0.5cm}
        
\begin{figure}[b]
    \centering
    \includegraphics[width=0.65\linewidth]{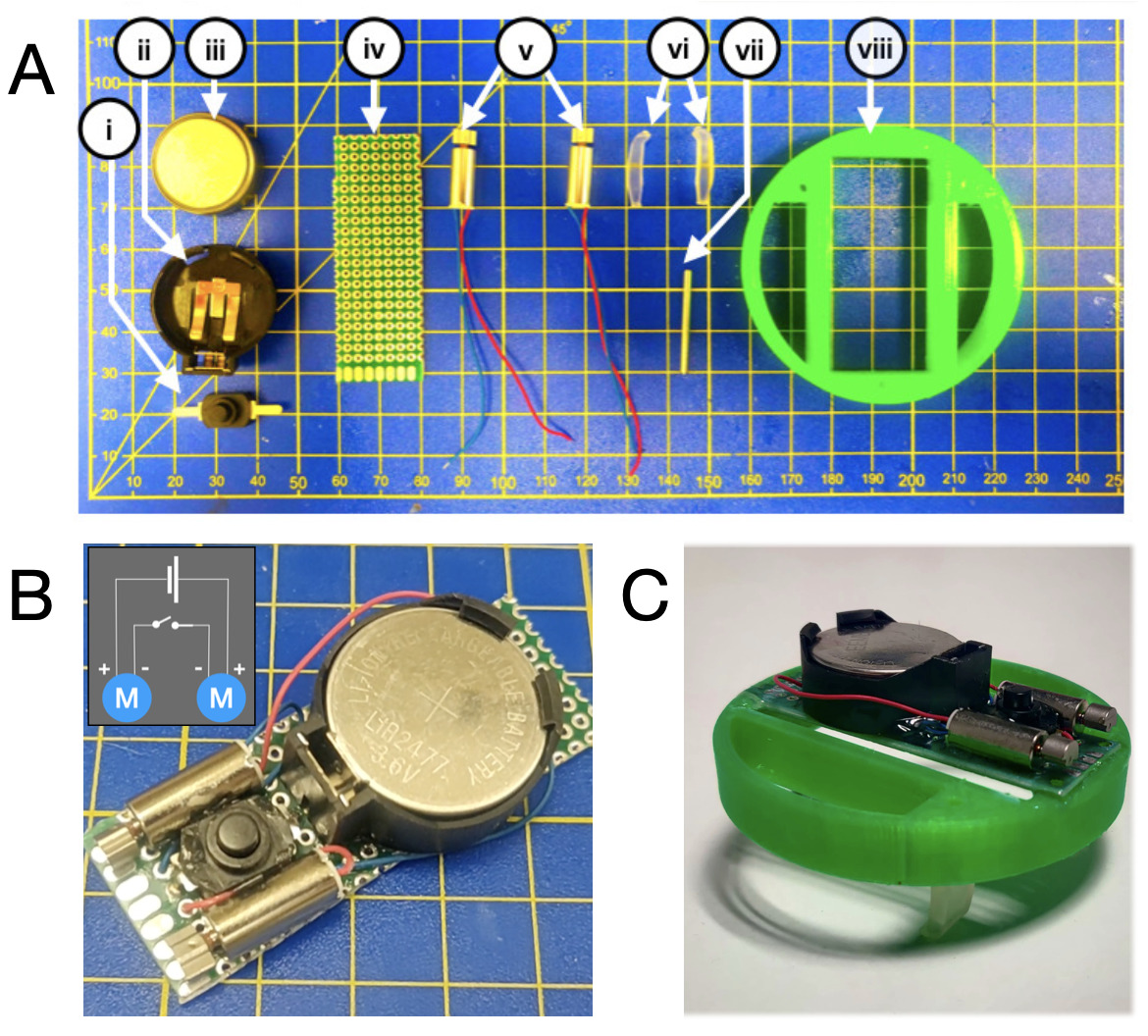}
        \caption{{\bf Assembly instructions for custom-built force-aligning robots.} (A) Assembly parts include: i - Switch; ii - battery house; iii - battery; iv - PCB board; v - vibration motors; vi - 3D printed soft elastic leg; vii - stiff leg; viii - 3D printed chassis (see Table~\ref{tblBoM} for further details). (B) Assembled electronic circuit (inset shows the circuit schematic). Note that motors are wired with opposite polarity for zero net torque. (C) A fully assembled robot with negative force-alignment.}
        \label{figBotAssembly}
\end{figure}

Custom-built robots were assembled in-house by combining the electronic circuit (presented in Fig.~\ref{figBotAssembly}) with 3D printed parts (files chassis.stl and flexiLeg.stl are in supporting materials), making a $\sim 6\;\rm{cm}$ robot. The bill of materials (BoM) is given in Table~\ref{tblBoM}. Following soldering and 3D printing, parts are glued together (estimated assembly time is 30 minutes).

    \begin{center}
        \def\arraystretch{1.5}
        \begin{table}
        \begin{tabular}{ |p{0.5cm}||p{2.65cm}|p{8cm}|p{0.5cm}|  }
             \hline
             \multicolumn{4}{|c|}{Bill of Materials} \\
             \hline
             \#  & Part Name & Ordering/Manufacturing Details  & Qty\\
             \hline
             i & Electric Switch &  12x8mm (LxW)
             
                                    DC 30V 1A Black On Off Mini Push Button Switch
                
                                    (1208-YD) &   1\\
             
             ii & Battery Holder &  29.1x26.4x15.8mm (LxWxH)
             
                                    CR2477 Black Coin Button Battery Holder   & 1\\
             
             iii & Battery &        24x7.7mm(DxH)
                
                                    LIR2477 Button Battery (3.6V, 200mAh) &   1\\
             
             iv & PCB &             20x50mm(WxL) 
             
                                    PCB board &  1\\
             
             v  & Vibration Motor & 6x14mm(DxL)
             
                                    DC Mini Vibration Motor 14000 RPM
                                    
                                    (BestTong A00000155)& 2 \\
             
             vi & Soft Legs &       20x7x5mm(LxWxT)
                                    
                                    3D Printer: Formlabs (Elastic50A Resin).
                                    
                                    CAD: flexileg.stl   &  2 \\
             
             vii & Stiff Leg &      2x25mm (DxL)
             
                                    stainless still pin  & 1 \\
             
             viii & Chassis  &      65x5mm (DxH)
             
                                    3D Printer: Prusa MK3 (PLA spool)
                                    
                                    CAD: chassis.stl  &  1 \\
             \hline
            
            \end{tabular}
            \caption{{\bf Bill of Materials for custom-built robots with controlled force alignment.} Part Number (\#) follows Fig.~\ref{figBotAssembly}A.} 
            
            \label{tblBoM}
        \end{table}    
    \end{center}

Custom-built robots were designed to maximize the ``restitution contrast'' between the soft and stiff legs. That is, making the soft legs very soft compared to the stiff leg. This effectively increases the ratio between $\tau_P/\tau_A$, increasing the magnitude of the curvity, $\kappa$ (see Section~\ref{secMechanics}). The soft leg material used is a proprietary material by Formlabs (Elastic 50A), a resin for SLA 3D printing that mimics the mechanical properties of (silicon rubber Young modulus of $\sim 50\; \rm{MPa}$), over 200 times softer than the stiff leg (steel $\sim 200\;\rm{GPa}$). The softness is also controlled by the geometry of the soft leg making a slandered body with a rectangular cross-section to best couple vertical vibrations from the vibration motors into a forward motion (see CAD design file flexileg.stl in extended materials). The natural vibration frequency of the soft legs is designed to match near resonance the vibration frequency of the motors ($\sim 200\; \rm{Hz}$). This was achieved by approximating the legs as a thin beam and using Euler-Bernoulli thin beam equations with a rectangular cross-section to maximize the backward-forward vibrations, and reduce the sideways and twisting vibrations.

\subsubsection{High Speed Imaging}\label{secHighSpeedImaging}

The rapid dynamics of the vibrating robots' contact with the floor were imaged using a high-speed camera (Phantom v2640). Robots were placed on a slope ($\sim 10^\circ$) facing perpendicular to the incline and imaged at 1000 frames per second (see Supporting Videos 3 and 4). The eccentric weights on the rotors of the vibration motors were observed to spin at $\sim 250\; \rm{Hz}$ (consistent with manufacturer specification of 14000 RPM). Contact with the floor was qualitatively estimated by observing the long periods of the stiff leg detaching from the ground, relative to the skidding motion of the soft legs on the ground which seem to slip but nearly not detach at all in the resolution used. The ratio between the acceleration induced by vibrations to the gravitational acceleration, $\gamma \equiv \frac{\omega^2 A}{g}$, can be estimated from Videos 3,4, where $\omega\approx 1600 rad/s$ is the vibration angular frequency, and $A \ge 1\; \rm{mm}$ is the oscillation amplitude resulting in $\gamma \approx 250$.  According to previous work~\cite{Vogel2011,Scholz2016}, at $\gamma > \sqrt{\pi ^2 +1}$, the bouncing dynamics are expected to be chaotic, as can be seen from the videos.

\subsubsection{Kilobot Based Robots}

\begin{figure}[h!]
    \centering
    \includegraphics[width=0.65\linewidth]{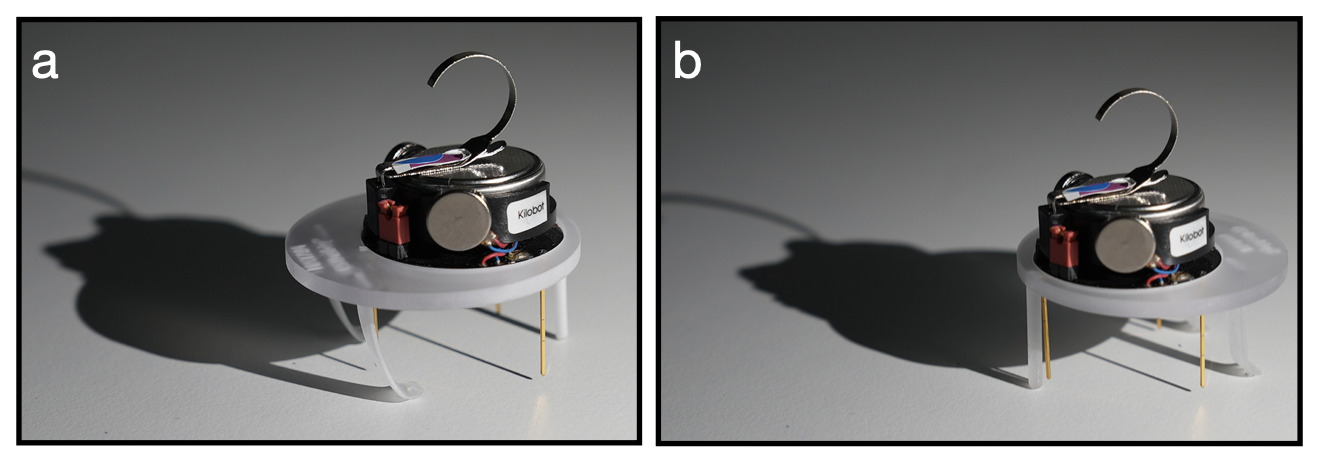}
        \caption{{\bf Commercial robots with 3D printed exoskeleton.} The position of the center of mass of commercial robots (Kilobot~\cite{Rubenstein2014}) was controlled using a 3D printed exoskeleton (see supporting files) using an exoskeleton where the center of mass is (a) behind the soft legs ($\delta<0$) or (b) in front of the soft legs ($\delta >0$).  Leading to $\kappa <0$ or $\kappa >0$ respectively (see Table~\ref{tblRobots}).}
        \label{figMorphobots}
\end{figure}

Robots were based on a commercially available multi-robot platform called Kilobtos~\cite{Rubenstein2014} modified with a 3D printed exoskeleton~\cite{Benzion2023}. With the exoskeleon, robots are $4.8\;\rm{cm}$ in diameter (see Fig.~\ref{figMorphobots}, with two exoskeleton designs used and supporting files fronter.stl and aligner.stl for negative and positive curvities respectively). The curvities of the robots were measured by tracking the trajectory and orientation of a robot (using trackPy~\cite{Allan2014}) while moving on an inclined plane. The orientation, $\theta$ follows a simple pendulum equation $\dot{\theta} = -\kappa \mu f \rm{sin}\theta$, from which the curvity was extracted in a process previously described in detail~\cite{Benzion2023}. The extracted curvities of the two designs were $\kappa \approx 0.3/b$ and $\kappa \approx -0.3/b$. An important benefit of working with the Kilobots is the ease of turning on and off a whole swarm of them (using a remote controller), facilitating experiments of collective effects.

\begin{center}
        \def\arraystretch{1.5}
        \begin{table}
        \begin{tabular}{ |p{3.5cm}||p{4.5cm}|p{2.5cm}|p{2.5cm}|}
            
             \hline
             Robot Type & Dimensions  (DxH) [mm] & Mass [g] & $\delta$ [mm]\\
             \hline
             Kilobot Fronter & 48x34 & 20 & $-13 \pm 2$\\
             \hline

             Kilobot Aligner & 48x34 & 20 & $20 \pm 2$\\
             \hline

             Custom Built Fronter & 65x37 & 34 & $-24 \pm 2$\\
             \hline

             Custom Built Aligner & 65x33 & 32 & $29 \pm 2$\\
             \hline

            \end{tabular}
            \caption{{\bf Robots Mechanical Properties.}} 
            
            \label{tblRobots}
        \end{table}    
    \end{center}

    \subsection{Cooperative Transport Experiments}\label{secSwarmExperimentalSetup}

        Experiments in cooperative transport were performed by monitoring the motion of a passive payload (polystyrene ring, see below) and a robotic swarm of Morphobots (Kilobots in an exoskeleton). In each experiment, the payload was symmetrically surrounded by idle robots, and then all the robots were set into motion simultaneously by setting both their motors at 130 drive intensity. The experiments tested the effect of the size of the swarm (between 1-53) and the size of the payload (7 cm - 32 cm).
        
\subsubsection{Swarm Experimental Setup}

\begin{figure}[b]
    \centering
    \includegraphics[width=0.65\linewidth]{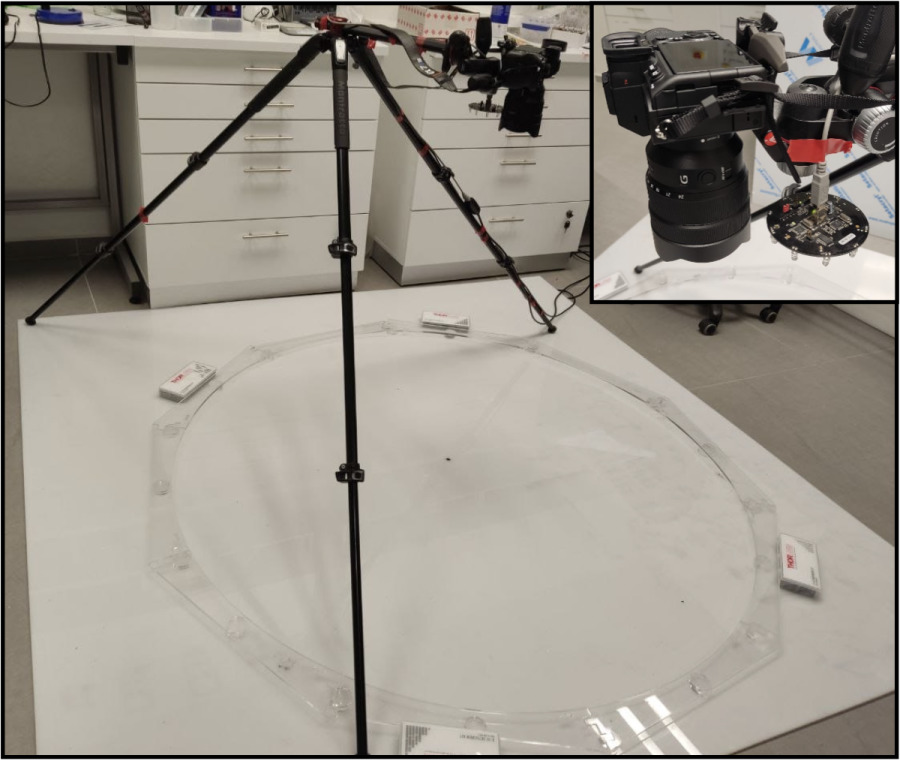}
    \caption{{\bf Experimental setup for cooperative transport experiments.} Arena diameter is 150 cm. Inset shows the camera and the swarms' overhead controller.}
    \label{figArena}
\end{figure}

Experiments were performed on a translucent acrylic plate with a laser-cut circular boundary of diameter 1.5 m. Dynamics were monitored by digital video recording using a Sony Alpha 7 s3 camera, on a tripod pointed down at the arena and mounted with an FE 12-24 mm f4g lens (see Fig.~\ref{figArena}). Videos were recorded at 25 frames per second at a 1920x1080 resolution for 20 minutes or until the payload reached the perimeter. The two-dimensional coordinates of the robots and payload, $\vec{R} = \left(x,y\right)$, were located using open CV2 circular Hough transform for identifying circles~\cite{Bradski2000}. Located particles were linked into trajectories $\vec{R}=\vec{R}\left(t\right)$ using trackPy~\cite{Allan2014}. From the trajectory of the payload, the mean square displacement (MSD) is calculated using
\begin{equation}
    \rm{MSD}\left(\Delta t\right) = 
        \langle \left| \vec{R}\left(\Delta t+t\right) - \vec{R}\left(t_i\right) \right| ^2 \rangle_t = \frac{1}{N}\sum_{i=1}^N\left|\vec{R}\left(\Delta t\right)-\vec{R}\left(t_i\right)\right| ^2.
        \label{eqMSD}
\end{equation}         
The power-law, $n$, of the mean square displacement of the trajectory of the payload of each experiment, was extracted from the slope of the linear fit of the MSD logarithm
\begin{equation}
    \rm{log}\left[(]\rm{MSD}\left(\Delta t\right)\right] = A + n \rm{log}\left(\Delta t\right).
        \label{eqPowerLaw}
\end{equation}
Below 1 minute, the MSD was typically still noisy given the locating resolution ($\sim 1 mm$) and trajectories typically were less than 10 minutes. Linear regression was employed between 1 minute and 5 minutes of the experiment, ensuring the power law is extracted above the noisier part of the locating and below the under-sampled portion. Reported values of exponents, $n$, in the text are the mean of 4 repetitions of swarm experiments with the error being the standard error of the mean.

\subsubsection{Experiments with different payload sizes}
\label{secPayloadSize}
\begin{figure}[h!]
    \centering
    \includegraphics[width=0.45\linewidth]{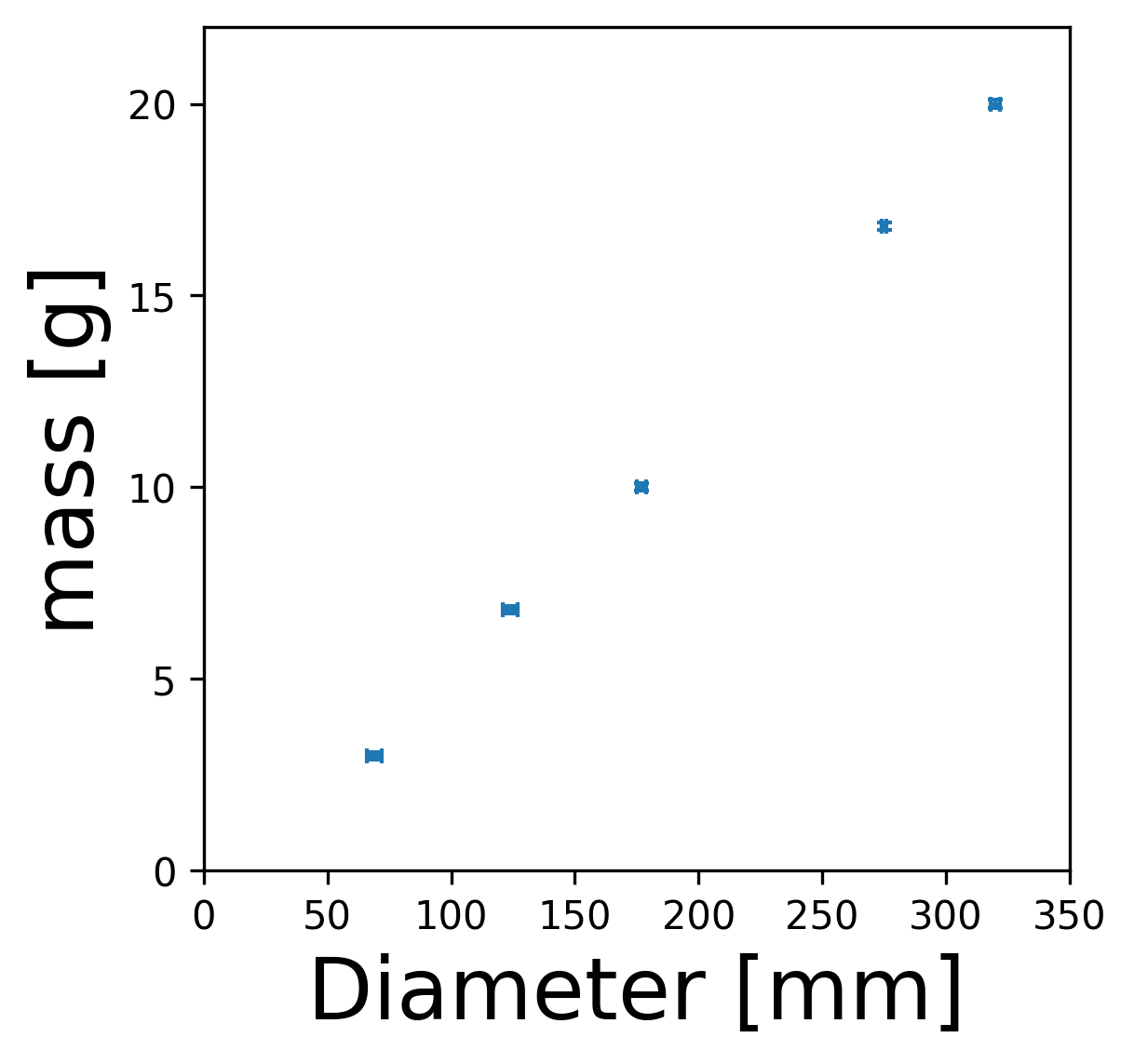}
    \caption{{\bf Linear increase of payload mass with diameter.}}
    \label{figPayloadMassDiameter}
 \end{figure}

Payload were made from polystyrene discs with mass kept linear with diameter by cutting an inner circle out of the disc, making rings of constant thickness (see Table~\ref{tblPayloads} and Fig.~\ref{figPayloadMassDiameter}). The static friction coefficient, $\mu_s$ was extracted from the angle of repose $\mu_s = \rm{tan}\phi$. The angle of repose was measured by placing the ring on an acrylic plate (same as the experimental arena), and tilting the plate while monitoring the tilt angle using a digital tilt meter ({\it Insize tilt meter} 2173-360). 

\begin{center}
    \def\arraystretch{1.5}
    \begin{table}[h!]
    \begin{tabular}{ |p{1cm}||p{2.5cm}|p{2.5cm}|p{2.5cm}|p{2.5cm}|p{2.5cm}|}
         \hline
         \multicolumn{6}{|c|}{Payloads} \\
         \hline
         Size & Diameter [mm] & Thickness  [mm] & Height [mm] & Mass [g] & $\mu_s$\\
         \hline
         7 & $68 \pm 3$ & (full) &  $39 \pm 1$ & $3 \pm 0.1$ & $0.47 \pm 0.05$\\

         12 & $124 \pm 3$ & $25 \pm 2$ &  $39 \pm 0.4$ & $6.8 \pm 0.1$ & $0.57 \pm 0.1$\\

         18 & $177 \pm 2$ & $24 \pm 1$ &  $40 \pm 0.4$ & $10 \pm 0.1$ & $0.48 \pm 0.03$\\
         
         28 & $275 \pm 1$ & $26 \pm 3$ &  $39 \pm 0.3$ & $16.8 \pm 0.1$ & $0.37 \pm 0.02$\\

         32 & $320 \pm 2$ & $29 \pm 4$ &  $39 \pm 0.9$ & $20 \pm 0.1$ & $0.33 \pm 0.01$\\
         
         \hline
        
        \end{tabular}
        \caption{{\bf Payloads properties}. Diameter, thickness, height, and static friction coefficient ($\mu_s$) are averages of 5 measurements with the error being standard deviation. $\mu_s$ is extracted from the angle of repose on a tilted acrylic surface.}        
        \label{tblPayloads}
    \end{table}    
\end{center}

\subsubsection{Measuring kissing number and effective attraction}

The effective attraction of the robots to the payload was characterized by the linear filling fraction of robots around the payload, $\lambda\equiv N_{\rm kiss}b/a$ (see Fig.~\ref{fig4} in the main text), where $N_{\rm kiss}$ is the number of robots touching the payload at a given moment. The kissing condition was defined when the distance from the located center of the robot, $\vec{r}$ to the located center of the payload, was less than the sum of radii plus a small buffer, $\epsilon = 3\;\rm{px}$, originating from the uncertainty in the payload's diameter ($\sim 2\;\rm{mm}$, see Table~\ref{tblPayloads}): $\left| \vec{R}-\vec{r}\right| < a+b+\epsilon$.

\subsubsection{Dependence on Swarm Size}\label{secSwarmSize}

\begin{figure}[h!]
    \centering                \includegraphics[width=0.5\linewidth]{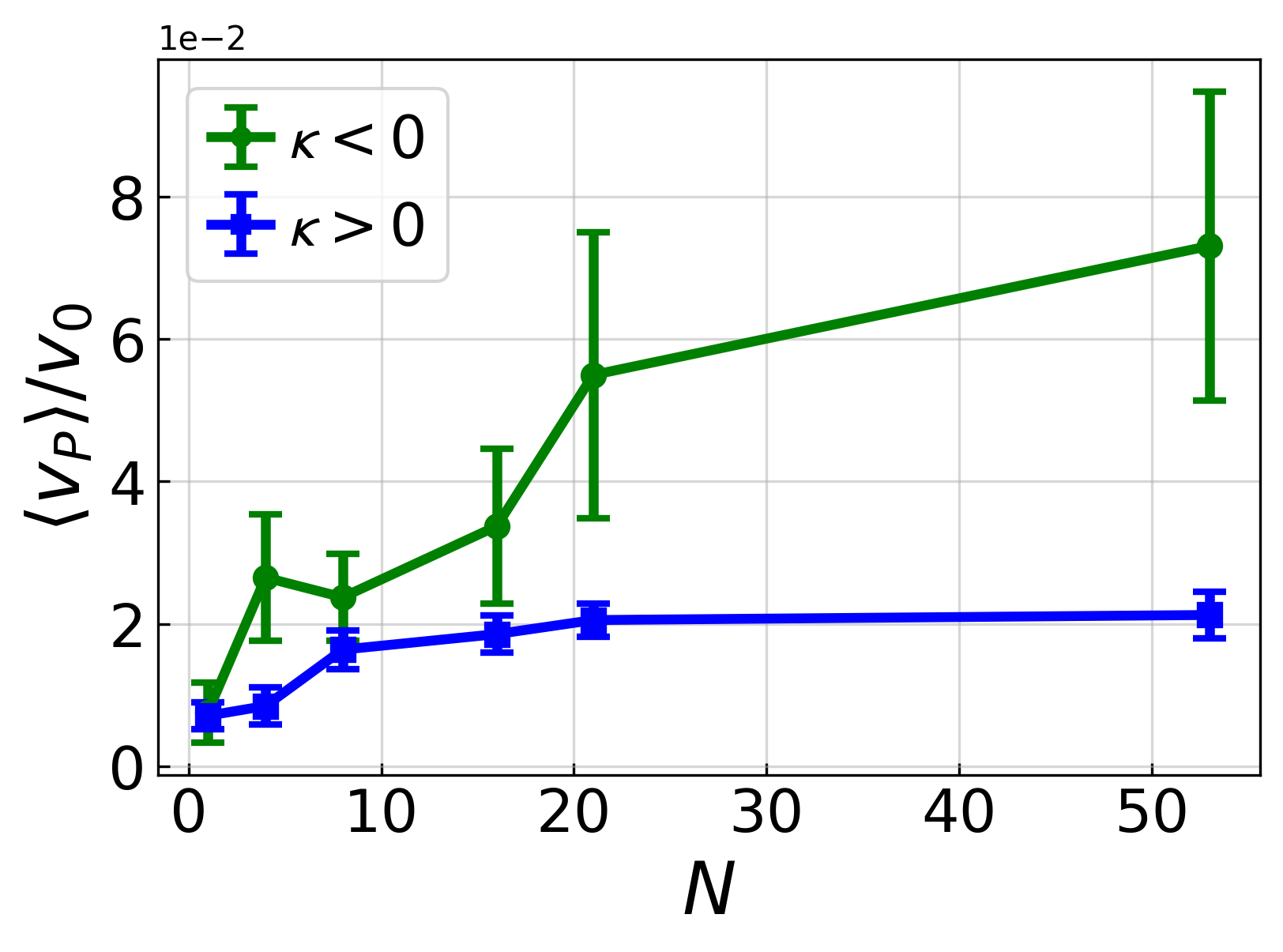}
    \caption{{\bf Cooperative transport increases with increasing swarm size for robots with negative curvity.} Measuring the mean speed of the payload ($\langle v_p \rangle$)  for swarms of different sizes ($N$), for both positive (blue) and negative (green) curivties, shows an increase in the effectiveness of transport with increasing swarm size.}
    \label{figSwarmSize}
 \end{figure}

The effect of the swarm size was measured using swarms ranging from 1 to 53 robots and a payload of 28 cm. Figure~\ref{figSwarmSize} shows the dependence of the mean speed of the payload ($v_p$) relative to the nominal speed of the robots ($v_0$) for swarms of different sizes. The mean payload speed was measured over 30-second windows. Each point is the average of 4 repetitions with error bars being standard error. The payload shows a consistent increase with swarm size for robots with negative curvity (green) but plateaus for robots with positive curvity (blue).

\section{Simulating Force Aligning Active Brownian Particles (FAABPs)}\label{secSimulations}

\subsection{Simulating individual FAABPs}

Trajectories of individual FAAPs (non-Brownian active particles such as Fig.~\ref{fig2}F in main text) were simulated using a 5th order Runge-Kutta numerical integrator (written in Python) for the time propagation of the following equations of motion:
\begin{eqnarray}		
&&\dot{\vec{r}} = v_0 \hat{e} + \mu \vec{f} \\
&&\dot{\hat{e}} = \kappa \left(\hat{e}\times\vec{v}\right)\times\hat{e},
\end{eqnarray}
with a constant external force pointing down the vertical direction ($\vec{f} = -f_0 \hat{y}$).

\subsection{Simulating interacting FAABPs with a passive particle}
The dynamical equations for each of the Force Aligning Active Brownian Particles (FAABP) model are
\begin{eqnarray}		
&&\dot{\vec{r}}_i = v^i_0 \hat{e}_i + \mu_i\vec{f}_i \\
&&\dot{\hat{e}}_i = \kappa \left(\hat{e}_i\times\vec{v}_i\right)\times\hat{e}_i + \vec{\xi}_i 
\end{eqnarray}
where $\xi_i$ is the magnitude of the orientational noise vector, with 
$\langle \xi_i^2\rangle = 2\Delta t k_BT$ ($\Delta t$ is the simulation time step, and $k_BT$ is the orientational noise as mentioned in the main text).
Each particle (active or passive) has a circular soft-core (of stiffness $S_0$), represented as a linear repulsive force with a cutoff. The force exerted on particle $i$ with radius $b$ by particle $j$ with radius $a$ is:

\begin{equation*}
    \vec{f}_{ij}\left(\vec{r}_{ij}\right) = 
        \begin{cases}
            S^j_0 \left(a+b-r_{ij}\right)\hat{r}_{ij}\; &{r_{ij} \le a+b}
            \\
            0 &{r_{ij} > a+b}
        \end{cases}
        \label{eqSimulation}
\end{equation*}

Mobility is inversely proportional to particle radius ($\mu_i \propto 1/a_i$), and the drag of larger payloads increases linearly with payload diameter (similar to Stokes law). The net force acting on particle $i$ is the summed contribution of the forces of all particles:

\begin{equation*}
    \vec{f}_i = \sum_{j\neq i} {\vec{f}_{ij}}
        \label{eqSimulationSum}
\end{equation*}
In the parameter range used ($b=0.05$, $v_0=30$, $k_BT\le 25000$, and $S_0 = 100$) particles do not overlap. The time step used in simulations was $dt\le 5\cdot 10^{-5}$ or $dt\le 1\cdot 10^{-5}$. The dynamics in Eqs.~\ref{eqSimulation} were simulated using a custom 5th Order Runge Kutta written in Python on a 13th Generation Intel i9 processor (3GHz) with 32 cores and 32 GB RAM. Each simulation starts by randomly placing the active particles around the passive particle on a torus (periodic boundary condition), monitoring their time evolution. Each run had between 1 and 1000 particles. The transport effect is also apparent in an open domain (non-periodic). Individual trajectories of passive particles were unwrapped to analyze their dynamic properties (e.g. payload's mean square displacement and mean speed).

\section{Derivation of the Mechanical Origin of Force Alignment}\label{secMechanics}

\begin{figure}[b]
    \centering                \includegraphics[width=1\linewidth]{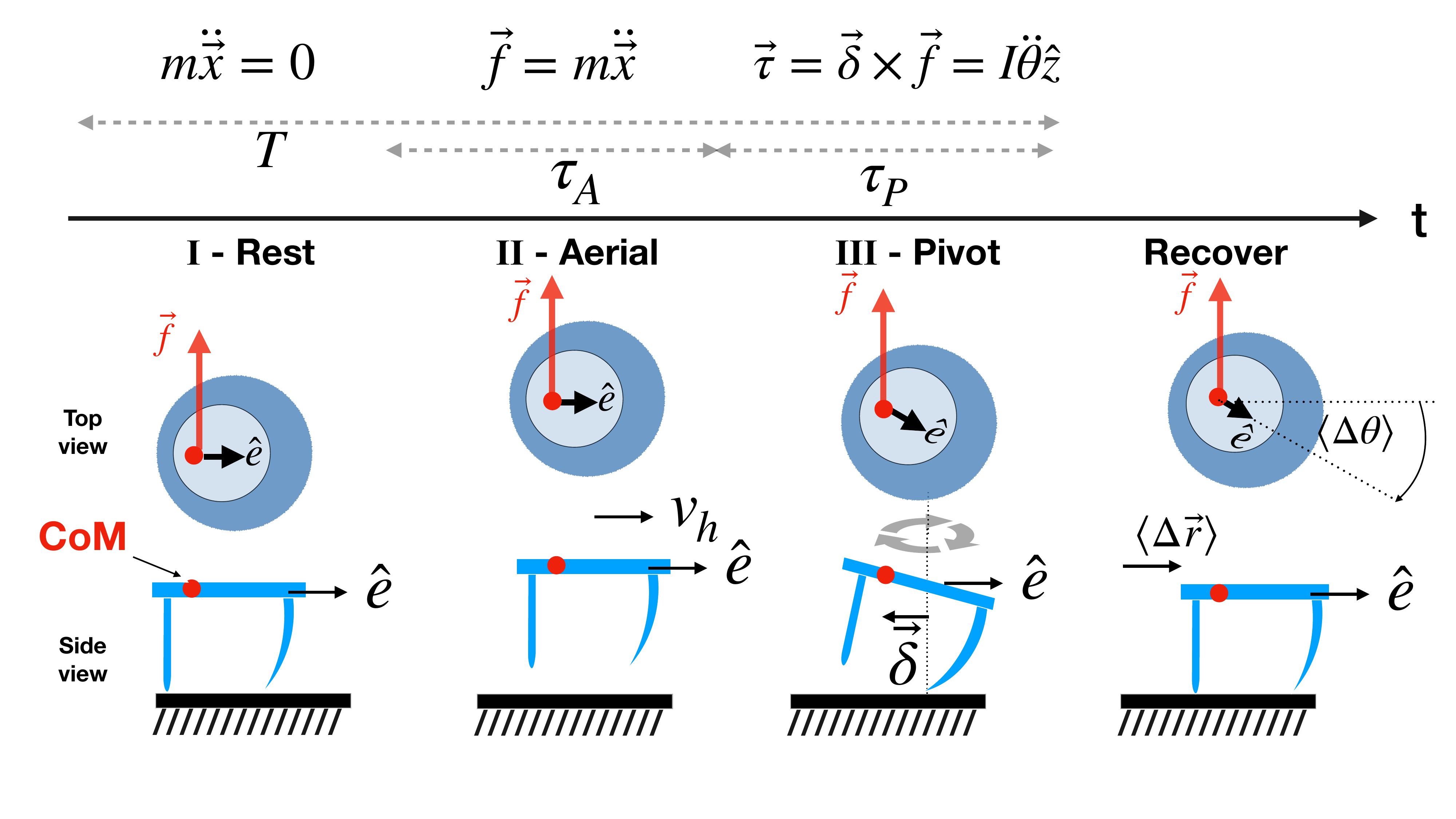}
    \caption{{\bf The rapid dynamics of hopping robots are divided into three phases I Rest, II Aerial, and III Pivot. Averaging over a period $T$ of the rapid dynamics, we find the mean displacement $\langle \Delta \vec{r}\rangle$ and rotation $\langle 
    \Delta \theta\rangle $ from which the effective equations of motion are derived.}}
    \label{figMechanicalModel}
 \end{figure}
 
We describe the quasi-2D motion of a motile robot using three phases (see Fig.~\ref{figMechanicalModel}). The motion is characterized by the linear and rotational accelerations ($\ddot{\vec{r}}$, $\ddot{\theta}$), which depend on the particle's mass $m$, which is by design displaced by an amount $\delta$ from its pivot axis that has the 2D moment of inertia $I$. The particle is subjected to an external body force $\vec{f}$ (acting in the plane). When the robot is touching the ground it loses all momentum, and has static friction with each leg that is touching the ground. Given that the robot vibrates at an order of 250 Hz, we will assume that the different phases happen rapidly, resulting in the slower net dynamics expanding to the first term in a Taylor series, the linear and rotational velocities. The three phases happen over a mean duration of $T$, over which each phase is averaged. Upon averaging the rapid dynamics, the robot will assume a mean displacement $\langle \Delta \vec{r}\rangle$, and a rotation, $\langle \Delta \theta \rangle$. Our goal is to find the mean linear velocity $\vec{v} \equiv \langle \Delta \vec{r}\rangle/T$ and mean rotational velocity $\vec{\omega} \equiv \langle\Delta \theta \rangle/T \hat{z} $ which define its effective equations of motion.

\subsection{Phases of motion of a bristle bot with asymmetric restitution}

\subsubsection{Resting Phase}
When all legs of the robot are touching the ground, both linear and rotational accelerations are zero, as the external force is balanced by the frictional force:
\begin{eqnarray}
&&\ddot{\vec{r}} = 0
\label{eqLinAcRest}\\
&&\ddot{\theta} = 0
\label{eqRotAcRest}
\end{eqnarray}

The robot does not have any initial linear or angular speeds, with mean displacement and rotation of:
\begin{eqnarray}
\label{eqLinVelRest}
&&\langle \Delta \vec{r}\rangle_R = 0\\
&& \langle \Delta \theta \rangle_R = 0
\label{eqRotVelRest}
\end{eqnarray}

\subsubsection{Aerial Phase}
In the aerial phase, the robot jumps with an initial speed along its heading $v_h\hat{e}$. The robot remains aloft for a duration $\tau_A$. Since it does not touch the ground, it only linearly accelerates by the external force 
\begin{eqnarray}
&&\vec{a} \equiv \ddot{\vec{r}} = \frac{1}{m}\vec{f}
\label{eqLinAcAerial}\\
&&\ddot{\theta} = 0.
\label{eqRotAcAerial}
\end{eqnarray}

The mean displacement and rotation during this phase are found through double integration of Eqs.~\ref{eqLinAcAerial},~\ref{eqRotAcAerial}, with the initial linear ($v_h\hat{e}$) and angular ($0$) velocities:

\begin{eqnarray}
&& \langle \Delta \vec{r}\rangle_A = v_h\tau_A \hat{e} + \frac{1}{2}\frac{\tau_A^2}{m} \vec{f} 
\label{eqLinDisAerial}\\
&& \langle \Delta \theta \rangle_A  = 0.
\label{eqRotDisAerial}
\end{eqnarray}

Upon averaging over the whole period ($T$), we arrive at Eq. \ref{eqForce} from the main text,
\begin{equation}
\dot{\vec{r}} \equiv \vec{v} = v_0 \hat{e} + \mu \vec{f},
\label{eqLinVelAerial}
\end{equation}
where the effective mobility is $\mu \equiv \frac{\tau_A^2}{2mT}$, and the self propulsion (``nominal'') speed is $v_0\equiv \frac{\tau_A}{T}v_h$.
In the aerial phase, the robot is torque-free and has no initial rotational velocity, so its mean rotational velocity is vanishing		
\begin{equation}
\vec{\omega}_A = 0.
\label{eqRotVelAerial}
\end{equation}

\subsubsection{Pivot Phase}
In the pivoting phase, the robot touches the ground with one leg and is subjected to static friction, defining a pivot axis. The external force now acts to rotate the robot with a lever arm $\vec{\delta}\equiv \delta\hat{e}$, and a moderate linear acceleration in the direction of the force
\begin{eqnarray}
&&\ddot{\vec{r}} \approx \frac{\delta^2}{I}\vec{f}
\label{eqLinAcPivot}\\
&&\vec{\alpha} \equiv \ddot{\theta} \hat{z} = \frac{\delta}{I}\hat{e}\times \vec{f}.
\label{eqRotAcPivot}
\end{eqnarray}

For simplicity, and without loss of generality we will neglect the linear displacement in the pivot phase as it is generally much smaller than in the linear phase ($\frac{\delta ^2}{I}\ll \frac{1}{m}$ ): the moment of inertia scales with the square of the robot radius ($b^2$) which in our design is much larger than $\delta^2$ (see Table~\ref{tblRobots}):
\begin{equation}
\langle \vec{r}\rangle_P \approx 0
\label{eqLinDispPivot}
\end{equation}

Similar to the mean linear translation in the aerial phase, the mean rotation in the pivot phase is given by double integration of the rotational acceleration (Eq.~\ref{eqRotAcPivot}), with zero initial rotational speed
\begin{eqnarray}
\langle \Delta \theta \rangle_P \hat{z} = \frac{1}{2}\frac{\delta \tau_P^2}{I}\hat{e}\times \vec{f}.
\label{eqRotDisAerial}
\end{eqnarray}

When averaging over the whole period ($T$), and re-expressing the force term using Eq.~\ref{eqLinVelAerial}, we get the mean angular speed 
\begin{equation}
\vec{\omega} =  \kappa \vec{v}\times\hat{e}.
\label{eqRotVelPivot}
\end{equation}

The microscopic parameters are absorbed in the resulting prefactor defined as $\kappa \equiv \frac{m\delta}{I}\left(\frac{\tau_P}{\tau_A}\right)^2$, which we call the curvity. Multiplying Eq. \ref{eqRotVelPivot} on both sides by $\hat{e}\times$ and recalling that $\hat{e}\times\hat{z} = -\hat{e}_\perp$, $\vec{\omega} = \dot \theta \hat{z}$, and that $\dot{\theta} \hat{e}_\perp=\dot{\hat{e}}$ we arrive at 
\begin{equation}
    \dot{\hat{e}} = \kappa \hat{e}\times\left(\hat{v}\times\hat{e}\right),
\label{eqRotVel}
\end{equation}
which is Eq. \ref{eqAlignment} in the main text.

\section{Derivation of the Geometrical Criterion for Cooperative Transport}~\label{secGeomCrit}
In this section, we present a detailed derivation of the geometrical origin of cooperative transport. We use the effective equations developed above from microscopic first principles, and show the role of geometry and the significance of the curvity of the active particles, $\kappa$, and the curvature of the passive payload, $1/a$, at triggering cooperative transport.

\subsection{Static Obstacle --- Dynamics of a FAABP in Fixed a Circular Potential}\label{secDynamics}

\subsubsection{Dynamical Equations of Motion}
The dynamical equations describing the time evolution of an active particle in a potential $U\left(\vec{r}\right)$, depend on three configurational degrees of freedom: the two-dimensional position, $\vec{r} = \left(x,y\right)$, and the heading, $\hat{e} = \left(\rm{cos}\theta,\rm{sin}\theta\right)$, where $\theta$ is the angle relative to the $x$ axis. The dynamical equations in vector form  are (Eq~\ref{eqForce},~\ref{eqAlignment} in the main text)
\begin{eqnarray}
&&\frac{d}{dt}\vec{r} \equiv \vec{v} = v_0 \hat{e} + \mu \vec{f} \\
&&\frac{d}{dt} \hat{e} = \kappa \hat{e} \times \left(\vec{v}\times\hat{e}\right).
\end{eqnarray}
The force on the active particle is a potential force, $\vec{f} =-\vec{\nabla}U\left(\vec{r}\right)$, and we will treat the circular payload as having a radially symmetric ($U=U(r)$), decaying ($U'\le 0$) potential. The force exerted by this potential on the active particle follows
$$ \vec{f} = -\vec{\nabla}U(r) = f(r)\hat{r}.$$
Independent of the explicit functional form of the decay (exponential, soft-core, etc.). The effective radius of the payload is such that the repulsive force at the radius exactly balances the self-propulsion of the active particle
$$ f(r=a) = v_0/\mu $$. 

We define $v_0\Gamma(r=a)$ to be the speed of a passive test particle of mobility $\mu$, and it is chosen to equal the active particle velocity at the perimeter of the payload (i.e. no motion $\dot{\vec{r}} = 0$)
\begin{equation}
\Gamma(r) \equiv \frac{\mu f(r)}{v_0},
\label{eqGamma}
\end{equation}
and the implicit radial force term becomes:
\begin{equation}
\vec{f}(r) = \frac{v_0}{\mu}\Gamma\left(r\right) \hat{r}.
\label{eqRadialForce}
\end{equation}
Plugging Eq.~\ref{eqRadialForce} into Eq.~\ref{eqForce} (main text) gives
\begin{equation}
\vec{v} = v_0\left(\hat{e} +\Gamma(r)\hat{r}\right),
\label{eqVelocityRadial}
\end{equation}
where $\vec{v}\equiv \frac{d}{dt}\vec{r}$ is the velocity of the active particle.
Plugging Eq.~\ref{eqVelocityRadial} into the alignment equation (Eq.~\ref{eqAlignment} in main text) gives
$$ \frac{d}{dt}\hat{e} = v_0\kappa \hat {e}\times\left[\left(\hat{e}+\Gamma(r)\hat{r}\right)\times\hat{e}\right] = v_0\kappa\Gamma(r) \hat{e}\times\left[\hat{r}\times\hat{e}\right].$$
The cross product between the radial unit vector and the orientation vector ($\hat{r}\times\hat{e}$) can be found by expressing the unit vectors in Cartesian basis where $\hat{r} = \rm{cos}\varphi \hat{x} + \rm{sin}\varphi \hat{y}$, and $\hat{e} = \rm{cos}\theta \hat{x} +\rm{sin}\theta \hat{y}$. The cross-product (square brackets on RHS) is 
$$\hat{r}\times\hat{e} = 	\begin{vmatrix} 
                    \hat{x} & \hat{y} & \hat{z} \\ 
                    \rm{cos}\varphi & \rm{sin}\varphi & 0 \\  
                    \rm{cos}\theta & \rm{sin}\theta & 0 \\  
                \end{vmatrix} 
            = \left[\rm{cos}\varphi\rm{sin}\theta - \rm{sin}\varphi\rm{cos}\theta \right] \hat{z}
            = \rm{sin}\left(\theta- \varphi\right)\hat{z} = \rm{sin}\psi \hat{z}.$$
            
We will denote $\psi$ as the difference between the angle of the unit orientation vector, and the azimuth of the active particle: $\psi \equiv \theta - \varphi$. Using $\psi$ will better reflect the rotational symmetry in the case of a stationary obstacle. This leads to the connection    
\begin{equation}
    \frac{d}{dt} \hat{e} = v_0\kappa\Gamma(r)\rm{sin}\psi\hat{e}\times\hat{z}.
\end{equation}
Recalling that the perpendicular orientation vector is equal to $\hat{e}_\perp = \hat{z}\times\hat{e}$, we find that the rate of change of the orientation vector 
\begin{equation}
    \frac{d}{dt}\hat{e} = -\kappa v_0 \Gamma(r) \rm{sin}\psi\hat{e}_\perp.
\end{equation} Using the identity for the rate of change of a unit vector $\frac{d}{dt}\hat{e} = \dot{\theta}\hat{e}_\perp$ we arrive at the dynamical change of the orientation coordinate:
\begin{equation}
    \dot\theta = -\kappa v_0\Gamma(r) \rm{sin}\psi. 
    \label{eqOrientation}
\end{equation}
We now turn to study the dynamical equation describing the position of the active particle $\vec{r}(t) = (r,\varphi)$, described by Eq. \ref{eqForce}. We start by expressing the orientation vector, $\hat{e}$, in polar coordinates: The radial component is given by $\left(\hat{e}\right)_r = \hat{e}\cdot \hat{r} = \left(\rm{cos}\theta \hat{x} +\rm{sin}\theta\hat{y}\right)\cdot \left(\rm{cos}\varphi \hat{x} +\rm{sin}\varphi\hat{y}\right)= \rm{cos}\psi$; the azimuthal component is $\left(\hat{e}\right)_\theta = \hat{e}\cdot\hat{\theta} = \left(\rm{cos}\theta\hat{x} + \rm{sin}\theta\hat{y}\right)\cdot \left(-\rm{sin}\varphi \hat{x} + \rm{cos}\varphi\hat{y}\right) = \rm{sin}\psi$. The orientation unit vector in polar coordinates is then
\begin{equation}
\hat{e} = \rm{cos}\psi \hat{r}+\rm{sin}\psi\hat{\varphi}.
\label{eqOrientationPolar}
\end{equation}

Expressing the velocity equation (Eq. \ref{eqForce}) in polar coordinates $\frac{d}{dt}\vec{r} = \dot{r}\hat{r} + r\dot{\theta}\hat{\theta}$, with the orientation expressed in polar coordinates (Eq. \ref{eqOrientationPolar}) gives the vector form of the velocity in polar coordinates:			
\begin{equation}
    \dot{r}\hat{r} + r\dot\varphi\hat\varphi = v_0\left(\rm{cos}\psi +\Gamma(r)\right)\hat{r} +v_0\rm{sin}\psi\hat{\varphi}.
\label{eqVelocityPolar}
\end{equation}

The azimuthal component is:
    \begin{equation}
     \dot\varphi = \frac{v_0}{r}\rm{sin}\psi\hat{\varphi}.
\label{eqAzimutalSpeed}
\end{equation}

Finally, we recall that $\dot\psi = \dot \theta - \dot\varphi$, and take the difference between the orientation equation (Eq. \ref{eqOrientation}) and the azimuthal component in Eq. \ref{eqAzimutalSpeed}, to arrive at the dynamical system describing the time evolution of the coordinates of FAABP in a fixed circular potential
\begin{eqnarray*}
&&\dot{r} =   v_0 \left[\rm{cos} \psi + \Gamma (r)\right] \\
&&\dot\psi = -v_0\left[\kappa \Gamma (r) +\frac{1}{r}\right] \rm{sin}\psi,
\end{eqnarray*}    
which are Eqs. \ref{eqRadial}, \ref{eqPsi} presented in the main text.

\subsubsection{Linear Stability Analysis of an Aligning Active Particle in a Circular Potential}\label{secLinearStability}
To find the fixed points of Eqs. \ref{eqRadial}, \ref{eqPsi}, we consider two cases:

\begin{enumerate}

\item $\kappa+1/a>0$\\   
There is a single fixed point at $\psi=\pi$, and $r=a$ which is a saddle point. An active particles approaches asymptotically to $\psi\rightarrow 0$ and  $r\rightarrow \infty$, effectively moving away from the obstacle.

\item	$\kappa+1/a<0$\\
There are three fixed points: a sink and two saddles. The sink is at $\psi=\pi$ and $r=a$, and two saddles can be found by solving the two transcendental equations 
\begin{eqnarray}
&&\tilde{u}(r) \equiv \kappa\Gamma(r)+1/r = 0\\
&&\rm{cos}\psi + \Gamma(r)=0,
\end{eqnarray}
with $r>a$. In Figure~\ref{fig5} in the main text, the phase portrait is given for $a=1$, $\kappa =-2$, with a repulsive profile of $\Gamma(r) = \rm{exp}\left[1-r/a\right]$, and the fixed points are found at $r \approx 2.68$, $\psi \approx \pi(1\pm 0.44)$.
\end{enumerate}		

\green{
\subsection{Cooperative transport as spontaneous symmetry breaking}
\subsubsection{Moving Obstacle --- Dynamical Equations of a FAABP in a Moving Circular Potential}\label{secMovingObject}
Cooperative transport in our model is triggered by spontaneous symmetry breaking, once a payload is moving. This can be illustrated by when the repulsive potential $U$ is moving in the lab frame where we find that FAABPs tend to continue to push it in the same direction of motion. We show this by considering a similar setting described above but with a moving potential: $U\left(\vec{r'}\left(t\right)\right)$, where $\vec{r'}$ is the center of a circular obstacle in the laboratory frame, with a potential repelling an active particle placed $\vec{r}$ {\it relative to the center of the obstacle} such that the force acting on the active particle is given by: $\vec{f} = \vec{f}\left(\vec{r'}+\vec{r}\right)$. The obstacle is assumed to move at a constant velocity along the $x$ direction (in the laboratory frame): $u\hat{x}$. Following similar steps to the derivation presented in Section ~\ref{secGeomCrit}, we arrive at a dynamical system with three equations of motion
\begin{eqnarray}
&&\dot{r} =   v_0 \left[\rm{cos} \psi + \Gamma (r)\right] - u\rm{cos}\varphi
\label{eqRadialMovingSI}\\
&&\dot\psi = -v_0\left[\kappa \Gamma (r) +\frac{1}{r}\right] \rm{sin}\psi - \frac{u}{r}\rm{sin}\varphi
\label{eqPsiMovingSI}\\
&&\dot\varphi = \frac{v_0}{r}\rm{sin}\psi + \frac{u}{r}\rm{sin}\varphi
\label{eqPhiMovingSI}
\end{eqnarray}
(Eqs.~\ref{eqRadialMoving},~\ref{eqPsiMoving},~\ref{eqPhiMoving} in the main text).

The first two equations (for $r$ and $\psi$) are similar to Eqs.~\ref{eqRadial} and \ref{eqPsi} in the main text, for the stationary payload with an added term ($\varphi$ dependent) stemming from its velocity. Eqs.~\ref{eqRadialMovingSI} and ~\ref{eqPsiMovingSI} have the same fixed points for the active particle to push against the repulsive potential ($\psi =\pi$ and $r\approx a$), provided that the azimuthal position is in front ($\varphi = 0$) or behind ($\varphi = \pi$) the direction of motion of the potential. Eq.~\ref{eqPhiMovingSI} reflects the added degree of freedom of the now-broken rotational symmetry. 

\subsubsection{Linear stability analysis of a FAABP pushing a moving obstacle}\label{secMovingObjectLinearStability}
For a pushing particle ($\psi=\pi$) Eq.~\ref{eqPhiMovingSI} (Eq.~\ref{eqPhiMoving} in the main text) has two fixed points, consistent with the $r$ and $\psi$ equations: an unstable fixed point for $\varphi =0$ and a stable fixed point for $\varphi = \pi$. The stable fixed point represents a force-aligning active particle that continues to push the payload in the direction it is already moving. This establishes the spontaneous symmetry breaking where a fluctuation in the velocity of the payload will perpetuate cooperative transport. Below we treat exclusively the pushing fixed point, showing its existence and stability.

We will consider the fixed point of Eqs.~\ref{eqRadialMovingSI}-~\ref{eqPhiMovingSI}, $\left(r*,\psi*,\varphi*\right)$, which satisfy $\dot{r}=0, \dot{\phi}=0, \dot{\psi}=0$, by: 
\begin{align}
    \psi^* &= \phi^* = \pi \\ 
     \Gamma(r*) &= 1 -\frac{u}{v_0}
\end{align}

Note that when the repulsive potential is moving, the fixed point need not be at $r=a$.
The Jacobian matrix of Eqs.~\ref{eqRadialMovingSI}-~\ref{eqPhiMovingSI} is:

\begin{align}
    \begin{bmatrix}
\frac{\partial \dot{r}}{\partial r} & \frac{\partial \dot{r}}{\partial \phi} & \frac{\partial \dot{r}}{\partial \psi} \\
\frac{\partial \dot{\phi}}{\partial r} & \frac{\partial \dot{\phi}}{\partial \phi} & \frac{\partial \dot{\phi}}{\partial \psi} \\
\frac{\partial \dot{\psi}}{\partial r} & \frac{\partial \dot{\psi}}{\partial \phi} & \frac{\partial \dot{\psi}}{\partial \psi}
\end{bmatrix} =
\begin{bmatrix}
v_0 \Gamma ' (r) 
& u\sin\phi
& -v_0 \sin \psi \\
-\frac{v_0}{r^2}\sin\psi-\frac{u}{r^2}\sin\phi& \frac{u}{r}\cos\phi
& \frac{v_0}{r} \cos \psi \\
-v_0\sin\psi(\kappa \Gamma '(r)-\frac{1}{r^2})+\frac{u}{r^2}\sin\phi & -\frac{u}{r}\cos\phi
& -v_0 \cos \psi (\kappa \Gamma (r) + \frac{1}{r})
    \end{bmatrix}
\end{align}

To ensure consistency across scales and to clarify intrinsic properties of the system, we make the Jacobian dimensionless by factoring out $\frac{v_0}{r}$.

\begin{align}
    \begin{bmatrix}
\frac{\partial \dot{r}}{\partial r} & \frac{\partial \dot{r}}{\partial \phi} & \frac{\partial \dot{r}}{\partial \psi} \\
\frac{\partial \dot{\phi}}{\partial r} & \frac{\partial \dot{\phi}}{\partial \phi} & \frac{\partial \dot{\phi}}{\partial \psi} \\
\frac{\partial \dot{\psi}}{\partial r} & \frac{\partial \dot{\psi}}{\partial \phi} & \frac{\partial \dot{\psi}}{\partial \psi}
\end{bmatrix} = \frac{v_0}{r}
\begin{bmatrix}
\Gamma ' (r) r & \frac{u r}{v_0}\sin\phi & - r \sin \psi\\
-\frac{1}{r}\sin\psi-\frac{u}{v_0 r}\sin\phi& \frac{u}{v_0}\cos\phi & \cos \psi \\
-\sin\psi(\kappa r \Gamma '(r)-\frac{1}{r})+\frac{u}{v_0 r}\sin\phi & -\frac{u}{v_0}\cos\phi
& - \cos \psi (\kappa r \Gamma (r) + 1)
    \end{bmatrix} 
\end{align}

At the fixed point $r^*$, $\phi^*=\pi$, $\psi^*=\pi$, we obtain:

\begin{align}
\frac{v_0}{r*}J = 
    \begin{bmatrix}
\Gamma'(r*)r* & 0 & 0 \\
0 & -\frac{u}{v_0} & -1 \\
0 & \frac{u}{v_0}
& \kappa r^* \Gamma(r*) + 1
    \end{bmatrix}
\end{align}

We will next explore the system's stability by considering if the  equation 
$$|J-\lambda I|=0$$
has all negative eigenvalues ($\lambda_i<0\forall i$) corresponding to a stable node. 

Filling the values from the Jacobian in and simplifying, we obtain:
$$\left[\Gamma'(r^*)r^* - \lambda\right]\left\{\left[- \frac{u}{v_0} - \lambda\right]\left[\kappa r^* \Gamma  (r*)+1-\lambda\right]+\frac{u}{v_0}\right\}=0.$$
The first eigenvalue is therefore:
$$\lambda_1 = \Gamma'(r^*)r^*$$
In the case of a decaying potential, $\Gamma'(r) < 0$ by definition, so also $\lambda_1 < 0$. Continuing to solve $\lambda_{2,3}$ gives:
$$\left[- \frac{u}{v_0} - \lambda\right]\left[\kappa r^* \Gamma  (r*)+1-\lambda\right]+\frac{u}{v_0}=0,$$
which simplifies to the quadratic equation for $\lambda$:
$$\lambda^2 + \lambda(\frac{u}{v_0} - \kappa r^* \Gamma  (r^*) -1) -\kappa r^* \Gamma  (r^*) \frac{u}{v_0}=0$$
\noindent Applying the following substitutions and simplifying gives:
\begin{align}
    \alpha &= \kappa r^* \Gamma  (r^*) + 1 \\
    \beta &= \frac{u}{v_0}
\end{align}
\begin{align}
    \lambda^2  + \left(\beta - \alpha \right)\lambda + \left(1-\alpha\right)\beta =0.
\end{align}
The roots of the  quadratic equation are
\begin{align}
    \lambda_{2,3} = \frac{(\alpha - \beta) \pm \sqrt{\left(\alpha-\beta\right)^2 +4\beta \left(\alpha -1\right)}}{2}
\end{align}
Stable nodes exist when both eigenvalues are real and negative.  The expression in the square root is positive when the payload is moving slower than the active particles $0<u/v_0<1$, and when the curvity is sufficiently negative $\kappa a < -1$. This is consistent with the empirical observations (both experimental and numerical). Note that here we focus on stable nodes, however even if the eigenvalue has an imaginary component, the fixed point can also be a stable spiral. The question of stability therefor rests on the sign of the real part of the eigenvalue. To find it we require both eigenvalues to be negative:
\begin{align}
    \lambda_2 &= (\alpha - \beta) - \sqrt{\left(\alpha-\beta\right)^2 +4\beta \left(\alpha -1\right)} < 0 \\
    \lambda_3 &= (\alpha - \beta) + \sqrt{\left(\alpha-\beta\right)^2 +4\beta \left(\alpha -1\right)} <  0
\end{align}
For the case of negative force alignment ($\kappa a <0$) and a slow moving payload ($u<v_0$) the first inequality is always satisfied and also $\lambda_2<0$. 
Lastly, we re-express the third eigenvalue,
\begin{align}
\lambda_3 = (\alpha - \beta)\left[1 + \sqrt{1 + 4\frac{\beta \left(\alpha -1\right)}{\left(\alpha - \beta\right)^2}}\right],
\end{align}
and see that the requirement for $\lambda_3<0$ is $\alpha<\beta$:
\begin{align}
\kappa r^*\Gamma\left(r^*\right)+1<\frac{u}{v_0},
\end{align}
which again is satisfied for a slow moving obstacle ($u<v_0$) and a sufficiently negative curvity ($\kappa a < -1$). In the limit of $u\rightarrow 0$, we recover the original condition for the isotropic system found (Eq.~\ref{eqTransport}).
}
\end{document}